\documentclass[12pt,a4paper]{article}
\pdfoutput=1
\usepackage{amssymb}
\usepackage{amsmath}
\usepackage{latexsym}
\usepackage[usenames]{color}
\usepackage{fancybox}
\usepackage{simplewick}
\usepackage{comment}
\usepackage{cite}
\usepackage{bm}
\usepackage{framed}
\definecolor{shadecolor}{rgb}{0.9,0.9,0.95}
\usepackage{setspace}
\usepackage[normalem]{ulem}
\usepackage[extdef]{delimset} % for brackets
\usepackage{multirow}
\usepackage[shortlabels]{enumitem}

\makeatletter \let\@keywords\@empty \let\@subject\@empty
\providecommand{\keywords}[1]{\gdef\@keywords{#1}}
\providecommand{\subject}[1]{\gdef\@subject{#1}}
\def\thetitle{\@title}
\def\theauthor{\@author}
\def\thesubject{\@subject}
\def\thedate{\@date}
\def\thekeywords{\@keywords}
\makeatother
\AtBeginDocument{%
\hypersetup{pdftitle={\thetitle}}%
\hypersetup{pdfauthor={\theauthor}}%
\hypersetup{pdfsubject={\thesubject}}%
\hypersetup{pdfkeywords={\thekeywords}}%
}
\makeatletter
\let\old@makecaption=\@makecaption
\def\@makecaption{\small\old@makecaption}
\makeatother
\providecommand{\hypersetup}[1]{}

\hypersetup{plainpages=false}
\hypersetup{pdfpagemode=UseNone}
\hypersetup{bookmarksnumbered=true}
\hypersetup{pdfstartview=FitH} % open with fit page
\hypersetup{colorlinks=false}
\hypersetup{citebordercolor={.5 1 .5}}
\hypersetup{urlbordercolor={.5 1 1}}
\hypersetup{linkbordercolor={1 .7 .7}}
\let\oldbfseries=\bfseries
\let\oldmdseries=\mdseries
\let\oldnormalfont=\normalfont
\renewcommand{\bfseries}{\oldbfseries\boldmath}
\renewcommand{\mdseries}{\oldmdseries\unboldmath}
\renewcommand{\normalfont}{\oldnormalfont\unboldmath}

\newcommand{\Left}{\mathrm{L}}
\newcommand{\Right}{\mathrm{R}}
\newcommand{\LeftRight}{\mathrm{L/R}}

\newcommand{\biloc}[2]{[#1|#2]}

\newcommand{\ham}{H}
\newcommand{\hamdens}{\mathcal{H}}
\newcommand{\mom}{P}
\newcommand{\oddchargedens}{\mathcal{P}}
\newcommand{\oddcharge}{P}
\newcommand{\momdens}{\mathcal{P}}
\newcommand{\genX}{X}
\newcommand{\charge}{Q}
\newcommand{\iom}{H}
\newcommand{\iomdens}{\mathcal{H}}
\newcommand{\chargedens}{\mathcal{Q}}
\newcommand{\loc}{A}
\newcommand{\current}{J}
\newcommand{\currentdens}{\mathcal{J}}
\newcommand{\currentdensI}{{\mathcal{J}}}
\newcommand{\currentdensG}{\mathcal{J}}
\newcommand{\Odens}{\mathcal{O}}

\newcommand{\alg}[1]{\mathfrak{#1}}

\newcommand{\order}[1]{\mathcal{O}(#1)}

\newcommand{\dd}{\mathrm{d}}
\newcommand{\ii}{\mathrm{i}}

\newcommand{\bdr}{\mathrm{bdr}}
\newcommand{\bulk}{\mathrm{bulk}}
\ifx\genfrac\sdflkaj

\else

\fi
\newcommand{\sfrac}[2]{{\textstyle\frac{#1}{#2}}}
\newcommand{\half}{\sfrac{1}{2}}
\newcommand{\ihalf}{\sfrac{i}{2}}
\newcommand{\quarter}{\sfrac{1}{4}}

\newcommand{\beq}{\begin{equation}}
\newcommand{\eeq}{\end{equation}}
\makeatletter
\def\mr@ignsp#1 {\ifx\:#1\@empty\else #1\expandafter\mr@ignsp\fi}%
\newcommand{\multiref}[1]{\begingroup%\let\protect\string%
\xdef\mr@no@sparg{\expandafter\mr@ignsp#1 \: }%
\def\mr@comma{}%
\@for\mr@refs:=\mr@no@sparg\do{\mr@comma\def\mr@comma{,}\ref{\mr@refs}}%
\endgroup}
\makeatother

\newcommand{\hypref}[2]{\ifx\href\asklfhas #2\else\href{#1}{#2}\fi}
\newcommand{\secref}[1]{Section~\multiref{#1}}
\newcommand{\appref}[1]{Appendix~\multiref{#1}}
\newcommand{\tabref}[1]{Table~\multiref{#1}}
\newcommand{\figref}[1]{Figure~\multiref{#1}}
\renewcommand{\eqref}[1]{(\multiref{#1})}

\makeatletter
\newlength{\apb@width}
\newcommand{\autoparbox}[2][c]{\settowidth{\apb@width}{#2}\parbox[#1]{\apb@width}{#2}}
\newcommand{\includegraphicsbox}[2][]{\autoparbox{\includegraphics[#1]{#2}}}
\makeatother

\usepackage{xcolor}
\definecolor{darkgreen}{rgb}{0,0.5,0}
\definecolor{darkblue}{cmyk}{0.9,0.9,0,0}
\definecolor{darkred}{rgb}{0.6,0,0.3}

\usepackage{graphicx}
\usepackage{tikz}
\usepackage[setpagesize=false,pagebackref=false, linktocpage, bookmarksopen=true, colorlinks=true, linkcolor=darkblue,citecolor=darkblue,urlcolor=black]{hyperref}
\usepackage{hyperref}

\newcommand{\id}{{\bf 1}}
\newcommand{\tr}{{\rm tr}}

\def\eqref#1{(\ref{#1})}

\def\beq{\begin{equation}}
\def\eeq{\end{equation}}

\newcommand{\rd}{\mathrm{d}}

\newcommand{\rB}{\mathrm{B}}

\newcommand{\rD}{\mathrm{D}}

\newcommand{\rL}{\mathrm{L}}

\newcommand{\rR}{\mathrm{R}}

\numberwithin{equation}{section}
\begin{document}
\thispagestyle{empty}

\setcounter{page}{1}
\setcounter{footnote}{0}
\setcounter{figure}{0}
%%%%%%%%%%%%%%%%%%%%%%%%%%%%%%%%%%%%%%%%%%%%%%%%%%%%%%%%%%%%%%%%%%%%%%%%%%%%%%%%%%%%%%%%%%%%%%%%%%%
\begin{flushright}
HU-EP-21/37
\end{flushright}

\thispagestyle{empty}

\begin{flushright}\footnotesize
%\texttt{arXiv:xxxx.xxxx}\\
\end{flushright}
%\vspace{1cm}

\begin{center}%
{\LARGE\textbf{\mathversion{bold}%
Irrelevant Deformations with\\ Boundaries and Defects
}\par}

\vspace{1cm}

 \textsc{
 Yunfeng Jiang$^{a,b}$,
 Florian Loebbert$^c$,
 De-liang Zhong$^d$}
 \vspace{8mm} \\
\textit{
$^{a}$Shing-Tung Yau Center and School of Physics, Southeast University,\\
 Nanjing 210096, China
}\\
\vspace{2mm}
\textit{$^{b}$CERN Theory Department, Geneva, Switzerland}
\\
\vspace{2mm}
\textit{%
$^c$ Institut f\"{u}r Physik, Humboldt-Universit\"{a}t zu Berlin, \\
Zum Gro{\ss}en Windkanal 6, 12489 Berlin, Germany
} \\
\vspace{2mm}
\textit{%
$^d$School of Physics and Astronomy, Tel Aviv University,\\
Ramat Aviv 69978, Israel
} \\
\vspace{2mm}
\texttt{jinagyf2008@gmail.com}
\\
\texttt{loebbert@physik.hu-berlin.de}
\\
\texttt{zdlzdlzdl@gmail.com}
\par\vspace{10mm}

\textbf{Abstract}
 \vspace{5mm}

\begin{minipage}{12cm}
We initiate the study of $T\bar T$-like irrelevant solvable deformations in quantum field theory with boundaries and defects. For this purpose, we employ a general formalism developed in the context of spin chains, which allows us to derive both, the deformed bulk and boundary/defect scattering matrices of integrable models. Using the deformed scattering matrices, we derive the flow equation for the deformed finite volume spectrum, as well as the cylinder partition function and the exact $g$-function.
\end{minipage}
\end{center}

%%%%%%%%%%%%%%%%%%%%%%%%%%%%%%%%%%%%%%%%%%%%%%%%%%%%%%%%%%%%%%%%%%%%%%%%%%%
%%%%%%%%%%%%%%%%%%%%%%%%%%%%%%%%%%%%%%%%%%%%%%%%%%%%%%%%%%%%%%%%%%%%%%%%%%%%

\newpage
\tableofcontents
\bigskip
\hrule
%%%%%%%%%%%%%%%%%%%%%%%%%%%%%%%%%%%%%%%%%%%%%%%%%%%%%%%%%%%%
%%%%%%%%%%%%%%%%%%%%%%%%%%%%%%%%%%%%%%%%%%%%%%%%%%%%%%%%%%%%%%%

%%%%%%%%%%%%%%%%%%%%%%%%%%%%%%%%%%%%%%%%%%%%%%%%%%%%%%%%%%%%%%%
\section{Introduction}
\label{sec:intro}
%%%%%%%%%%%%%%%%%%%%%%%%%%%%%%%%%%%%%%%%%%%%%%%%%%%%%%%%%%%%%%
The $T\bar{T}$ deformation \cite{Smirnov:2016lqw,Cavaglia:2016oda} has extended our understanding of quantum field theory (QFT) in 1+1 dimensions. The deformed theory has a number of remarkable features which are quite robust and universal. Although a better understanding of certain seemingly pathological properties, such as the Hagedorn behavior of the density of states and the complex energy spectrum is required, it is reasonable to suspect that such deformations are meaningful and lead to an interesting generalization of the usual local QFTs.\par

Local QFTs can have extended structures such as boundaries and defects. They are interesting for several reasons. First, they describe real physical situations where boundaries and defects are ubiquitous. Second, from a more formal point of view, the defects contain a lot of useful information on the bulk theory, see for example \cite{Cardy:1986ie,Oshikawa:1996ww,Frohlich:2004ef,Quella:2006de,Chang:2018iay}.
A natural question is then how such non-local structures are modified under $T\bar{T}$ deformation? The goal of the current work is to initiate investigations into this direction.\par

We start with a special situation where both the local QFT and the boundary/defect are integrable. The reason is that the $T\bar{T}$ deformation preserves integrability and as a result, such theories can be studied by the powerful toolkit of integrable models, in particular employing the scattering picture and factorized S-matrices. Such a description is very convenient for the study of the $T\bar{T}$ deformed integrable QFTs because the bulk S-matrix is deformed in a simple way \cite{Smirnov:2016lqw,Cavaglia:2016oda,Dubovsky:2017cnj}. The new ingredients in the boundary/defect case are the deformed boundary and defect S-matrices, which will be derived along the lines of \cite{Bargheer:2008jt,Bargheer:2009xy,Loebbert:2012yd}. Once the deformed S-matrices are known, we can apply the machinery of integrability to compute various important physical quantities. This procedure is universal and does not depend on details of the theory under consideration. In this sense, the deformed quantum model is more straightforward to study than the classical counterpart.\par

Another important motivation for the study of the $T\bar{T}$ deformation comes from the theory of integrable models. It is by now known that for integrable models and CFTs, the $T\bar{T}$ deformation is a special case of a more general family of integrable deformations triggered by bilinear operators. Such integrable bilinear deformations lead to a novel type of integrable models. For relativistic integrable QFT and CFT, more general solvable bilinear deformations with an additional $U(1)$ current have been studied in \cite{Guica:2017lia,Guica:2019vnb,Anous:2019osb,Nakayama:2018ujt,Frolov:2019xzi,Aguilera-Damia:2019tpe,Chakraborty:2019mdf,Apolo:2018qpq}. Deformations  involving higher conserved currents have been proposed and studied in \cite{LeFloch:2019rut,LeFloch:2019wlf,Conti:2019dxg,Hernandez-Chifflet:2019sua,Doyon:2021tzy}. For spin chains, similar deformations have been introduced even before the $T\bar T$ deformations in \cite{Bargheer:2008jt,Bargheer:2009xy} and are called bilocal deformations, \emph{cf.}\ \tabref{tab:DifferentDeformations}.
\begin{table}[t]
\renewcommand{\arraystretch}{1.2}
\begin{center}
\begin{tabular}{|l||c|c|}
\hline
Deformation& Field Theory& Spin Chain
\\\hline
 $T\bar T$ & Bilinear ($\mom$ and $\ham$) & $\times$
\\
 Generalized $T\bar T$ &Bilinear ($\charge_r$ and $\charge_s$) $\stackrel{\text{map}}{\longleftrightarrow}$\hspace{-6mm} & \hspace{2.5mm}Bilocal ($\charge_r$ and $\charge_s$)
\\
Boost&$\times$&Bilocal ($\id$ and $\charge_r$)
\\\hline
\end{tabular}
\caption{Different types of deformations for field theory and lattice models with the respective charges employed for their construction. For integrable models, the momentum $\mom=\charge_1$ and Hamiltonian $\ham=\charge_2$ form part of a larger set of conserved charges $\charge_r$. Boost-type deformations can be understood as bilocal deformations formed from the identity operator $\id$ and a charge $\charge_r$. While it is currently not known how to define the original $T\bar T$ deformation for the spin chain case, the so-called boost deformations have only been defined for the lattice model.}
\label{tab:DifferentDeformations}
\end{center}
\end{table}
 Their relation to $T\bar{T}$-like deformations was first pointed out in \cite{Pozsgay:2019ekd,Marchetto:2019yyt}. Also in the spin chain case deformations using `internal' symmetries have been explored \cite{Beisert:2013voa} and, importantly for the present paper, a generalization to systems with open boundaries exists \cite{Loebbert:2012yd}. Due to the discrete nature of the spin chain model, it is yet unknown how to define the `real' $T\bar{T}$ deformation for lattice models, which requires the conserved momentum current. Bilinear deformations of integrable non-relativistic quantum many-body systems such as the 1d Bose gas (non-linear Schr\"odinger model) have been studied in \cite{Cardy:2020olv,Jiang:2020nnb,Hansen:2020hrs,Ceschin:2020jto,Chen:2020jdi,Esper:2021hfq}. For these models, the $T\bar{T}$ deformation can be defined and it was found that the deformation has the effect of changing the length of fundamental particles of the model. The simplest deformation that changes the length of particles is the hard rod deformation defined in \cite{Cardy:2020olv,Jiang:2020nnb}. It turns out that the hard rod deformations can be defined for a wide range of models including lattice models. The hard rod deformation of spin chain models can be identified with the constrained XXZ \cite{Alcaraz1999,Karnaukhov2001,Alcaraz2007} and folded XXZ spin chain \cite{Zadnik2020a,Zadnik2020b,Pozsgay:2021dyy,Gombor:2021nhn,Pozsgay:2021rwc}, which have recently received renewed interest from different perspectives \cite{Borla:2019chl,Yang:2019mft,Moudgalya:2021xlu}.\par

Integrable boundaries and defects have played an important role for integrable models. Therefore, it is of great interest to also study these novel types of integrable models with boundaries and defects. The bilocal deformation of quantum spin chains in the presence of integrable boundary conditions was first considered by one of the authors in \cite{Loebbert:2012yd}, where the deformed reflection matrix has been derived generalizing the bulk approach of \cite{Bargheer:2008jt,Bargheer:2009xy}. The method turns out to be general and can be applied to other types of integrable models including relativistic QFTs. In the scattering picture, integrable boundaries and defects are uniquely characterized by their scattering amplitudes with the particles. Therefore, our strategy is to determine the deformed boundary and defect scattering amplitudes, which can be achieved by a natural generalization of the method in \cite{Loebbert:2012yd}. Together with the deformed bulk S-matrix, we can compute important physical quantities in the deformed theory. We consider two such quantities, which are the deformed spectrum and the exact $g$-function (or Affleck-Ludwig boundary entropy).\par

The rest of the paper is structured as follows. In \secref{sec:IQFTs}, we give a brief review of boundary and defect integrable QFTs. We perform a classical analysis of the deformed Lagrangian in \secref{sec:ClassDef}. In \secref{sec:GeneralDefs}, we give a detailed discussion of integrable bilocal deformations, which applies to general integrable systems including spin chains, relativistic and non-relativistic IQFTs. We then derive the deformed boundary and defect amplitudes in \secref{sec:OpenChannelScatteringDefs}. Together with the deformed bulk S-matrix, we determine the deformed finite volume spectrum and the exact $g$-function in \secref{sec:spectrum} and \secref{sec:TorusA}, respectively. We conclude in \secref{sec:conclude} and discuss future directions. More details are given in the three appendices.

%%%%%%%%%%%%%%%%%%%%%%%%%%%%%%%%%%%%%%%%%%%%%%%%%%%%%%%%%%%%%%
\section{Boundary and Defect Quantum Field Theory}
\label{sec:IQFTs}
%%%%%%%%%%%%%%%%%%%%%%%%%%%%%%%%%%%%%%%%%%%%%%%%%%%%%%%%%%%%%%

In this section, we review some general properties of boundary integrable quantum field theories (IQFTs) following the seminal paper of Ghoshal and Zamolodchikov \cite{Ghoshal:1993tm}. We consider two-dimensional Euclidean quantum field theories in flat spacetime with Cartesian coordinates $(x^1,x^2)=(x,y)$. To quantize the system one has to choose the direction of time.  We consider the theory defined on $(x,y) \in (s_\Left,s_\Right) \times \mathbb{R}$, where $s_\Left$ and $s_\Right$ denote the generic positions of the left and right boundary, respectively. In the so-called \textit{open channel}, one chooses the $y$ direction as the direction of time. In this channel, the Hamiltonian reads
\begin{equation} \label{eq:HamiltonianGeneric}
\iom_r=\int_{s_\Left}^{s_\Right}\iomdens_{r}(\phi(x,y))\,\dd x-\theta_{r\Left}(\phi(s_L,y))+\theta_{r\Right}(\phi(s_L,y)),
%=P_r+\bar P_r+\theta_{2r-}(t)+\theta_{2r+}(t),
\end{equation}
with the left/right boundary function (or boundary Hamiltonian) $\theta_{r\LeftRight}(t)=\theta_{r}(s_\LeftRight,t)$.
Here, we denote the ``fundamental" field by $\phi(x,y)$, and we assume that the Hamiltonian is a local function of $\phi$ and its derivatives $\partial^\mu \phi$. We could also have some boundary degrees of freedom governed by the boundary function $\theta_{r\LeftRight}$ living on the $x=s_{\LeftRight}$ boundary line, which is a function of $\phi(x=s_{\LeftRight},y)$ and its time derivatives (recall the time direction is $y$).

It is worth mentioning that the Hamiltonian \eqref{eq:HamiltonianGeneric} does not have the most general form of a bulk-boundary interaction.
Following Ghoshal and Zamolodchikov \cite{Ghoshal:1993tm} we make the following assumptions:
\begin{itemize}
    \item We consider a single scalar field in the bulk.
    \item There are no new boundary degrees of freedom, but the boundary function depends only on the boundary field, which is identical to the bulk field evaluated at the boundary.
    \item The boundary Hamiltonian $\theta$ is of \textit{potential} type, \emph{i.e.}\  it is only a function of the boundary field, but not of its derivatives.
\end{itemize}
For the rest of the paper, unless otherwise stated, we will make these assumptions for the Hamiltonian, and we will also sometimes call the boundary function $\theta$ the \emph{boundary potential}.

\subsection{Conserved Charges}
\label{sec:ConservedCharges}

We will be interested in studying deformations generated by conserved charges. Hence, in particular in the context of integrability the space of deformations is very rich.

%%%%%%%%%%%%%

\paragraph{Bulk Charges.}

Let us first briefly recall how integrability is realized when there are no boundaries.
 It is convenient to introduce complex coordinates
 \footnote{The corresponding metric is off-diagonal, whose non-vanishing components are $\eta_{z\bar{z}} = \eta_{\bar{z}z} = 1/2$, and $\eta^{z\bar{z}} = \eta^{\bar{z}z} = 2$.}
\begin{equation}
z=x+i y,
\qquad\qquad
\bar z=x-iy,
\end{equation}
such that
\begin{equation}
\partial_z=\half(\partial_x-i\partial_y),
\qquad\qquad
\partial_{\bar z}=\half(\partial_x+i\partial_y).
\end{equation}
Assuming Lorentz invariance, the conserved charges associated with the spacetime symmetries are fully encoded in the stress-energy tensor. Let us introduce some notation for the components of the stress tensor in complex coordinates:
\begin{equation}
T=-T_{zz},
\qquad
\bar T=- T_{\bar z \bar z},
\qquad
\Theta=\bar \Theta=T_{z\bar z}=T_{\bar z z}.
\end{equation}
Here $T_{\mu \nu}$ is the stress tensor of the theory.

In addition to the charges associated with the stress tensor, integrable theories possess an infinite set of mutually commuting integrals of motion, which can be understood as higher spin generalizations of the stress tensor charges. Conventionally, one can construct those conserved charges in terms of some local spin-$s$ fields $T_s$ and $\Theta_s$, which satisfy
\begin{equation}
\partial_{\bar z} T_{s+1}=\partial_z \Theta_{s-1}.
\label{eq:Thetas}
\end{equation}
Here the allowed spins take values in a subset of integers and are fully determined by the theory.
Assuming parity invariance, operators with negative spin $s$ are related to the operators with positive spin by parity, so it is convenient to define the barred charges as
\beq
\bar{T}_{s+1} = \Theta_{-s -1}, \qquad \bar{\Theta}_{s-1} = T_{-s+1}.
\eeq
With this definition, the conservation equations can now be summarized as
\begin{equation} \label{eqn-BulkChargeConser}
\partial_{\bar z} T_{s+1}=\partial_z \Theta_{s-1}.
\qquad\qquad
\partial_z\bar T_{s+1}=\partial_{\bar z}\bar\Theta_{s-1},
\end{equation}
where the spin label $s$ is assumed to be non-negative. We shall employ this convention in the following.

Using the local fields, we can construct conserved charges by means of contour integrals. Defining
\begin{align}
    I_s&=\int_\mathcal{C}\ \brk*{\dd z\ T_{s+1}+\dd\bar{z}\ \Theta_{s-1}},
    &
    \bar I_s&=\int_\mathcal{C}\ \brk*{\dd \bar z\ \bar T_{s+1}+\dd z\ \bar \Theta_{s-1}},
\end{align}
one immediately sees that $I_s$ and $\bar I_s$ are independent of the choice of the contour because of \eqref{eqn-BulkChargeConser}.
One recovers the charges in Euclidean coordinates by taking linear combinations of $I_s$ and $\bar I_s$. For instance, the momentum and Hamiltonian are expressed as
\begin{equation}
\mom=-\ii \brk1{I_1-\bar I_1},
\qquad\qquad
\ham=I_1+\bar I_1.
\end{equation}
Similarly one obtains higher $\ham$- and $\mom$-type charges:
\begin{equation}
\ham_s=I_s+\bar I_s =\int \hamdens_s(x) \dd x,
\qquad\qquad
\mom_s=-\ii \brk*{I_s-\bar I_s} =\int \momdens_s(x) \dd x.
\label{eq:higherHP}
\end{equation}
Here the local densities $\hamdens_s(x)$ and $\momdens_s(x)$ are defined as
\begin{align}
\iomdens_{s}
&=T_{s+1}+\Theta_{s-1}+\bar T_{s+1}+\bar \Theta_{s-1},
\\
\oddchargedens_{s}
&=-\ii \brk*{T_{s+1}+\Theta_{s-1}-\bar T_{s+1}-\bar \Theta_{s-1}}.
\end{align}
Conservation of the above charges follows from current conservation
\begin{equation}
\partial_y\iomdens_s=-\partial_x\currentdensI_{\iomdens_s},
\qquad\qquad
\partial_y\oddchargedens_s=-\partial_x\currentdensG_{\momdens_s},
\end{equation}
%%%%%%%
where the generalized current densities take the form
\begin{align}
\currentdensI_{\iomdens_s}
&=-\ii\brk*{T_{s+1}-\Theta_{s-1}-\bar T_{s+1}+\bar \Theta_{s-1}},
\\
\currentdensG_{\momdens_s}
&=-T_{s+1}+\Theta_{s-1}-\bar T_{s+1}+\bar \Theta_{s-1}.
\end{align}
Note that the $\mom$-type charges are only conserved in the bulk, while conserved $\ham$-type charges can also be defined in the boundary model as described in the following paragraph.
It will also be useful to employ the universal notation
\begin{equation}
\charge_{2s}=\iom_{2s-1},
\qquad\qquad
\charge_{2s-1}=\oddcharge_{2s-1},
\qquad
s=1,2,\dots.
\label{eq:EvenOddCharges}
\end{equation}
Similarly we define the shifted current densities as
\begin{equation}
\currentdens_{2s}=\currentdensI_{\iomdens_{2s-1}},
\qquad\qquad
\currentdens_{2s-1}=\currentdensG_{\momdens_{2s-1}},
\qquad
s=1,2,\dots,
\end{equation}
such that the conservation equation takes the form
\begin{equation}
\partial_y \chargedens_{s}=-\partial_x \currentdens_{s}.
\label{eq:ConsEq}
\end{equation}
Thus, for the boundary model the even charges $\charge_{2s}$ are conserved, while the odd charges $\charge_{2s+1}$ will only be conserved in the bulk.
The Heisenberg equation for the charges, which will be used in some of the following derivations, reads
\begin{equation}
\partial_y\chargedens_r=\comm{\ham}{\chargedens_r}.
\label{eq:HeisEq}
\end{equation}
Finally, the one-particle eigenvalues of the different types of charges are denoted according to
\begin{equation}
\charge_s\to q_s(u),
\qquad
\ham_s \to e_s(u),
\qquad
\mom_s \to p_s(u),
\end{equation}
where the eigenvalues of the $\charge_s$ split up as
\begin{equation}
q_{2s}(u)=e_{2s-1}(u),
\qquad
q_{2s-1}(u)=p_{2s-1}(u).
\end{equation}

%%%%%%%%%%%%%

\paragraph*{Boundary Effects.}

In the presence of a boundary, the bulk conservation law \eqref{eqn-BulkChargeConser} is not sufficient to guarantee the conservation of $I_s, \bar I_s$. For notational simplicity, we consider a system with only one boundary and we set $s_{\rL} = - \infty, s_{\rR} = 0$.
Let us compare the conserved charges at different times. Consider say $I_1$ defined as an integral over two different contours $\mathcal{C}_1$ and $\mathcal{C}_2$, which are parallel to the $x$-axis but end on different points $y_1, y_2$ on the $y$ axis. Taking the difference we find
\begin{equation}
\begin{aligned}
I_1^{\mathcal{C}_1} - I_1^{\mathcal{C}_2} &= \ii \int_{y_1}^{y_2} \rd y\ (T - \Theta)\Big|_{x=0} \neq 0, \\
\bar I_1^{\mathcal{C}_1} - \bar I_1^{\mathcal{C}_2} &= \ii \int_{y_1}^{y_2} \rd y\ (\bar \Theta - \bar T)\Big|_{x=0} \neq 0, \\
\end{aligned}
\end{equation}
for generic $T_s, \Theta_s, y_1, y_2$, \emph{i.e.}\ the charges are a priori not conserved. A simple fix is to choose an appropriate boundary function $\theta$ as given in \eqref{eq:HamiltonianGeneric}, such that
\begin{equation}
    T_{xy}\big|_{x=0} = -\ii(T-\bar T)\big|_{x=0} = \frac{\rd}{\rd y} \theta(y),
\end{equation}
where $\theta(y)=\theta(x=0,y)$ is some local boundary field. Physically, the original conformal boundary condition $T_{xy} =0 $ just means that there is no energy/momentum flow passing through the impenetrable boundary at $x=0$. The generalized boundary condition with $T_{xy} \neq 0$ then implies that the energy flow can be absorbed by a ``potential term'' on the boundary.
Adding $\theta(y)$, the new bulk-boundary Hamiltonian of the form
\begin{equation}
H = \int_{-\infty}^0 \rd x\ (T+\bar T + \Theta + \bar \Theta) + \theta(y)
\end{equation}
is a conserved quantity since the additional $\theta$-term compensates the boundary contribution.

Similarly, we can introduce generalized boundary conditions for higher spin fields,
\begin{equation} \label{eqn-BdrCondHigherCharge}
-\ii(T_{r+1}-\bar T_{r+1} +\bar \Theta_{r-1} - \Theta_{r-1})\Big|_{x=0} = \frac{\rd}{\rd y} \theta_r(y),
\end{equation}
and the higher conserved charges are now given by
\begin{equation}
    H_r = \int_{-\infty}^0 \rd x\ (T_{r+1}+\bar T_{r+1} +\bar \Theta_{r-1} + \Theta_{r-1}) + \theta_r(y).
\end{equation}

The discussion above immediately generalizes to the case with two boundaries: one only needs to impose the boundary condition for both, the left ($\Left$) and right ($\Right$) boundary. Conservation of the charges requires that the boundary field obeys
\begin{equation}
\partial_y \theta_{r\LeftRight}(t)
=
 \currentdensI_{\iomdens_r}(s_\LeftRight).
\label{eq:opencurrentboundfield}
\end{equation}

To summarize, we can construct conserved charges in the presence of the boundaries by adding a boundary function $\theta_r$ to the Hamiltonian for each boundary. At the same time, only particular ($\ham$-type, see \eqref{eq:higherHP}) linear combinations of the two types of conserved charges $I_s$ and $\bar I_s $ are conserved. The odd $P$-type charges are not conserved in the presence of boundaries since translation symmetry and its higher spin generalizations are broken by the boundaries.

\subsection{Open- and Closed-Channel Picture}
\label{sec:openclosed}

\begin{figure}[t]
\begin{center}
\includegraphics[scale=0.45]{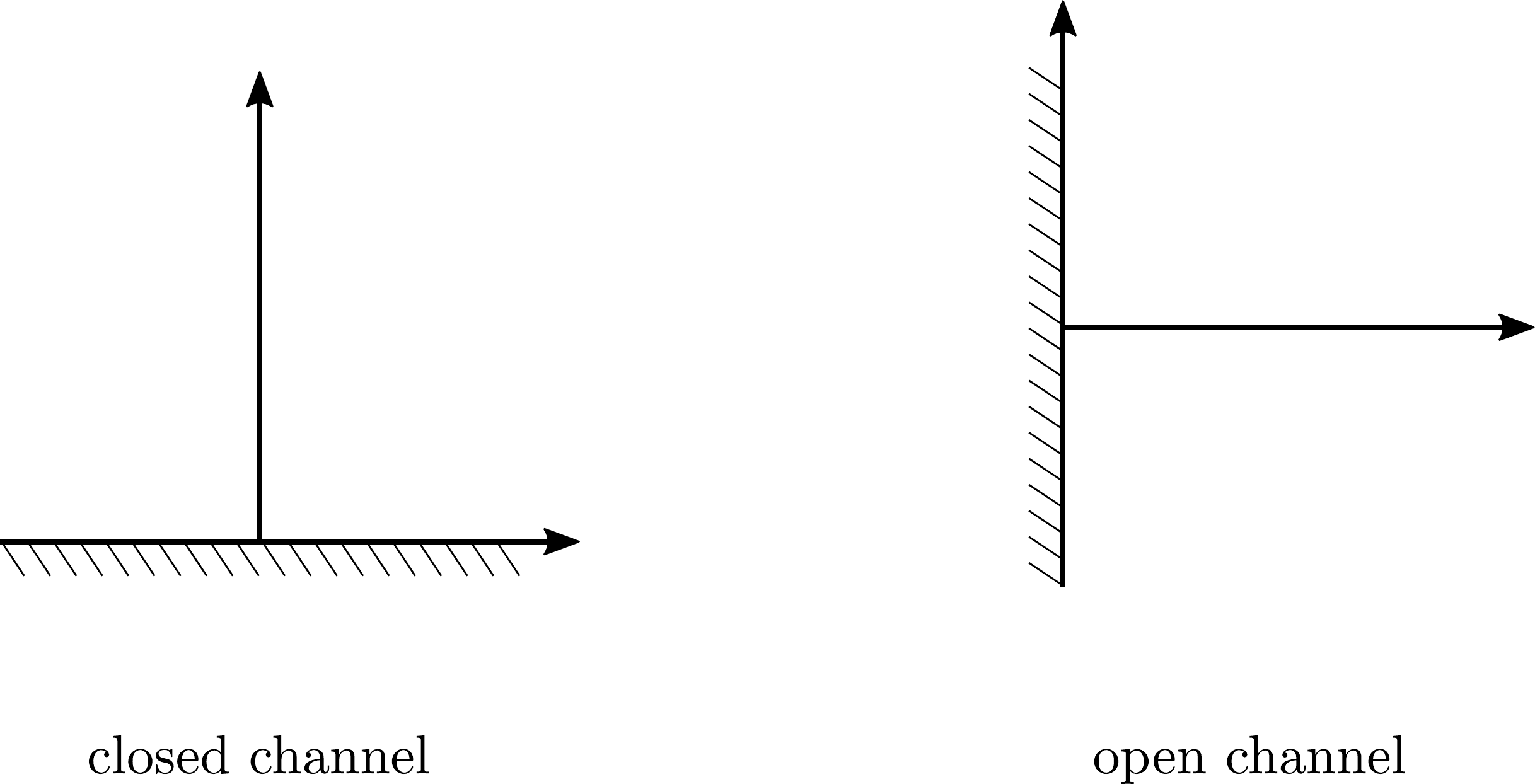}
\caption{Different channels for boundary QFTs. The left and right panels are the closed and open channels, respectively.}
\label{fig:channel}
\end{center}
\end{figure}

The boundary can be placed either in the spatial or temporal direction as is shown in \figref{fig:channel}. These choices give different but equivalent descriptions of the same theory.
In the open channel, the boundary is placed in the spatial direction. Without loss of generality, we can put it
at $x=0$ and define the system on the left half line, as is shown in the right panel of \figref{fig:channel}. The Hamiltonian in the open channel reads
\begin{align}
H_{\text{open}}=\int_{-\infty}^0 \dd x\, T_{yy}(x) \, .
\end{align}
Correlation functions in this channel are computed by
\begin{align}
\langle O_1(x_1,y_1)\ldots O_N(x_N,y_N)\rangle=\frac{{_\rB}\langle 0|\mathcal{T}_y[O_1(x_1,y_1)\ldots O_N(x_N,y_N)]|0\rangle_{\rB}}
{{_\rB}\langle 0|0\rangle_{\rB}}
\end{align}
where
$|0\rangle_{\rB}$ is the ground state of $H_{\text{open}}$ and
\begin{align}
O_i(x,y)=e^{-y H_{\text{open}}}O_i(x,0) e^{y H_{\text{open}}} \, .
\end{align}
Here, $\mathcal{T}_y$ means ordering with respect to the $y$ direction.\par

In the closed channel, the boundary is placed in the Euclidean time direction, see the left panel in \figref{fig:channel}. The Hamiltonian is the same as for a QFT without boundaries and given by
\begin{align}
H_{\text{closed}}=\int_{-\infty}^{\infty} \dd y \, T_{xx}(y),
\end{align}
where $T_{yy}(x)$ is one of the components of the stress energy tensor $T_{\mu\nu}(x)$ on some constant $x$ slice.\par

Since the boundary is placed in the temporal direction, it should be understood as a \emph{boundary state}, which we denote by $|B\rangle$. Correlation functions of local operators in the closed channel with one boundary at $x=0$ are given by
\begin{align}
\langle O(x_1,y_1)\ldots O(x_N,y_N)\rangle=\frac{\langle 0|\mathcal{T}_x[O(x_1,y_1)\ldots O(x_N,y_N)]|B\rangle}{\langle 0|B\rangle},
\end{align}
where $|0\rangle$ is the ground state of $H_{\text{closed}}$ and $\mathcal{T}_x$ is the time ordering. We have
\begin{align}
O_i(x,y)=e^{x H_{\text{closed}}}O_i(0,y)e^{-x H_{\text{closed}}}\, .
\end{align}

%%%%%%%%%%%%%
\subsection{Scattering Picture}

In this section, we will discuss how the usual scattering picture arises in integrable theories. Deformations of the respective bulk and boundary scattering matrices will then be discussed in the subsequent sections.

Consider a multi-particle scattering process in two dimensions. Each particle is specified by its energy and momentum, which satisfies the relativistic dispersion relation $e^2 - p^2 = m^2$, where $m$ is the mass of the particle. It is convenient to parametrize the energy $e$ and momentum $p$ by the \textit{rapidity} variable $u$, defined via $(e,p) = (m\cosh u, m\sinh u)$. We thus describe scattering processes in terms of the rapidities of the particles.

\paragraph{Bulk Scattering.}

Integrability imposes strong constraints on a multi-particle scattering process. All multi-particle processes factorize into consecutive two-to-two scattering events. The compatibility condition is that the order of this factorization does not affect the physical amplitudes, which leads to the Yang-Baxter equation
\begin{equation}
    S_{12} S_{13} S_{23} = S_{23} S_{13} S_{12},
\end{equation}
where $S_{ij} = S_{ij}(u_i,u_j)$ denotes the S-matrix of the $ij$ two-to-two scattering process. Here $u_j$ represents the rapidity of the scattered particle $j$.

\paragraph{Boundary Scattering.}

In the presence of boundaries, we have to add new integredients to the Yang-Baxter equation to retain integrability.
Let us first recall the ordinary quantum mechanical scattering picture in the presence of boundaries. Consider an incoming plane wave moving towards the boundary. If the boundary is impenetrable, this wave must be fully reflected.
For a unitary theory with a single type of particle, the reflection amplitude can only differ by a phase from the incoming amplitude. For a system with more than one particle type, the reflection matrix is a unitary matrix. This \textit{boundary scattering matrix} or \textit{boundary scattering phase}, respectively, will be denoted by $S_\Left(u)$ or $S_\Right(u)$. Here  $u$ stands for the rapidity of the reflected particle and $\Left$ or $\Right$ denotes the left or right boundary, respectively.

Integrability imposes non-trivial constraints on the boundary S-matrix. Consider a two-particle scattering process, where the two particles scatter through each other in the bulk, hit the boundary and then return into the bulk. Demanding that this process is independent of the order of scattering, we obtain the \textit{boundary Yang-Baxter equation} for \emph{e.g.}\ the left boundary scattering matrix $S_\Left$:
\begin{equation}
    S_\Left(u_2) S_{21}(u_1+u_2) S_\Left(u_1) S_{12}(u_1-u_2) = S_{12}(u_1-u_2)S_\Left(u_1) S_{21}(u_1+u_2) S_\Left(u_2).
\end{equation}
In addition to that, the boundary S-matrix must also satisfy the crossing and unitarity constraints \cite{Ghoshal:1993tm}. The boundary S-matrices can be obtained non-perturbatively by solving these constraints.

%%%%%

\paragraph{Faddeev-Zamolodchikov Algebra.}

Before discussing the more general cases, let us first review a convenient way of describing integrable scattering processes. Consider the asymptotic states of an integrable field theory, which can be expressed in terms of creation operators $A^\dagger(u)$ via
\begin{equation} \label{eqn-inoutstate}
|u_1,\cdots, u_n \rangle_\textrm{in/out} = A^\dagger(u_1) \cdots A^\dagger(u_n) |0 \rangle.
\end{equation}
The in/out states are distinguished by the relative ordering of the particle rapidities: if $u_1 >u_2 > \cdots u_n$ it is understood to be an ``in-state''; if instead the rapidities are ordered as $u_1<u_2 < \cdots <u_n$ it is understood as an ``out-state''. It is worth to mentioning that the creation operators are \textit{not} the creation operators of the free theory, instead they take into account all interactions.
However, a nice feature of those creation operators is that we can describe the scattering process as in a free theory, where the creation operators satisfy
\begin{equation}
    A^\dagger(u_1) A^\dagger(u_2) = S(u_1,u_2) A^\dagger(u_2) A^\dagger(u_1).
\end{equation}
Here the coefficient $S$ represents the two-to-two scattering matrix. Since asymptotic states diagonalize the local charges, we can write
\begin{equation}
\comm{I_s}{ A^\dagger(u)} = \gamma^{(s)} e^{su} A^\dagger(u),
\qquad
\comm{\bar I_s}{A^\dagger(u)} = \gamma^{(s)} e^{-su} A^\dagger(u),
\end{equation}
where the $\gamma^{(s)}$ are constants determined by the theory.

%%%%%%%%%%%%

\paragraph{Defect Theory.}

A defect is a generalization of an impenetrable boundary.
Quantum mechanically, the physics is identical to the scattering off a potential barrier, where we are allowed to have both, transmitted and reflected waves. For a unitary theory, the sum of the modulus of the reflection and transmission amplitude is $1$. The study of integrable line defects was initiated in \cite{Delfino:1994nr,Delfino:1994nx}.

In integrable theories it is convenient to describe integrable defects by the Faddeev-Zamolodchikov (FZ) algebra. The reason is that the generalized FZ algebra in the presence of a defect has the same structure as the usual asymptotic quantum mechanical scattering picture.

The line defect separates the space into two parts, which will be called the left and right part. We denote the FZ operators in the two parts by $A^{\dagger}(u)$ and $B^{\dagger}(u)$, respectively, and we assume that the action of the defect can be described by a defect creating operator
$\mathbf{D}^{\dagger}$. The most general defect algebra is then
\begin{align}
A^{\dagger}(u)\mathbf{D}^{\dagger}=&\,R(u) A^{\dagger}(-u)\mathbf{D}^{\dagger}+T_-(u)\mathbf{D}^{\dagger}A^{\dagger}(u),\\\nonumber
\mathbf{D}^{\dagger}B^{\dagger}(u)=&\,R(u)\mathbf{D}^{\dagger}B^{\dagger}(-u)+T_+(u)A^{\dagger}(u)\mathbf{D}^{\dagger}.
\end{align}
The physical meaning of these equations is quite clear: for instance, the first equation describes the scattering process of the left particle off the defect.

It has been proven that the only integrable defects in an interacting theory are topological ones \cite{Castro-Alvaredo:2002qcm}. These defects are purely transmissive. Setting the reflection coefficient to zero ($R(u)=0$) in the previous equations, the algebra satisfied by these operators reads
\begin{align}
A^{\dagger}(u)\mathbf{D}^{\dagger}=T_-(u)\,\mathbf{D}^{\dagger}B^{\dagger}(u),\qquad
\mathbf{D}^{\dagger}B^{\dagger}(-u)=T_+(-u)A^{\dagger}(-u)\mathbf{D}^{\dagger},
\end{align}
where $T_{\pm}(u)$ are the transition amplitudes. For a parity symmetric theory, we have $T_-(u)=T_+(u)$. The asymptotic states are given by
\begin{align}
|u_1,\cdots,u_M;v_1,\cdots,v_N\rangle\equiv A^{\dagger}(u_1)\cdots A^{\dagger}(u_M)\mathbf{D}^{\dagger}
B^{\dagger}(v_1)\cdots B^{\dagger}(v_N)|0\rangle \, .
\end{align}
Here, we have introduced an evident ket notation $|u;v\rangle$ to denote the rapidities of the left/right side of space

One can have non-topological defects, but the price to pay is that the theory has to be free, namely the bulk S-matrices are simply $S= \pm 1$. In this case, the creation operators $A^\dagger, B^\dagger $ are identical, and the most general asymptotic state can be written as
\begin{align}
    |u\rangle_{\rD}=a(u)|u;\varnothing\rangle+b(u)|\varnothing;u\rangle+c(u)|-u;\varnothing\rangle,
    \end{align}
where we have
\begin{align}
    T(u)=\frac{b(u)}{a(u)},\qquad R(u)=\frac{c(u)}{a(u)}.
\end{align}
This completes our introduction to two-dimensional quantum field theories whose deformations will be discussed in the following.
%%%%%%%%%%%%

%%%%%%%%%%%%%%%%%%%%%%%%%%%%%%%%%%%%%%%%%%%%%%%%%%%%%%%

%%%%%%%%%%%%%%%%%%%%%%%%%%%%%%%%%%%%%%%%%%%%%%%%%%%%%%%%%%%%%%
\section{Classical Analysis}
\label{sec:ClassDef}

The purpose of this section is to perform a classical analysis of deformations of the above family of boundary field theories. In particular, we wish to understand for which boundary conditions the $T\bar T$-deformed theories allow for integrability. We will see that in order to preserve integrability at leading order in the deformation parameter, the boundary potential of the undeformed theory has to be zero. More explicitly, we will take a given bulk model and evaluate the constraints on the boundary conditions, which arise from constructing a first higher spin charge. We will first briefly discuss the free scalar  as an illustrative example and then study the $T\bar T$-deformed model. Details on the deformed Sine-Gordon model are given in \appref{apd-SG}.

%%%%%%%%%
\subsection{Lagrangian Description}

In order to study classical deformations it will be convenient to use the Lagrangian description. Recall that our field theory is defined on the $-x$ axis, by taking $s_\Left = - \infty, s_\Right = 0$, with the fields vanishing at $x = s_\Left$.
Performing a Legendre transformation of the Hamiltonian \eqref{eq:HamiltonianGeneric}, the action of the system can be written as
\beq \label{eqn-Lag-General}
S(\phi)  = \int_{-\infty}^\infty \rd y \int_{-\infty}^0 \rd x\, \mathcal{L}(\phi, \partial_\mu \phi)  + \int dy\, \mathcal{L}_\bdr(\phi_\bdr) \, ,
\eeq
where $\phi_\bdr(y) = \phi(x=0,y)$.
Here, the boundary Lagrangian $\mathcal{L}_\bdr$ is
essentially the boundary potential $\theta_{2\Right}(x=0,y)$ defined in \eqref{eq:HamiltonianGeneric}.

As usual, the classical equation of motion can be obtained by taking the functional variation. In the bulk, the equation of motion is given by the usual
Euler-Lagrange equation, which does not depend on the boundary Lagrangian.
On the contrary, the equation of motion for the boundary field $\phi_\bdr$ depends on both, the bulk and the boundary Lagrangian. The dependence on the bulk Lagrangian comes from the total spatial derivative terms in the bulk (recall that these are $x$-derivatives in our setup). For instance, suppose under a field variation $\phi \rightarrow \phi + \delta \phi$ the variation of the bulk Lagrangian contains a spatial derivative term
$\frac{\partial \mathcal{L}}{\partial (\partial_x \phi)} \partial_x \delta \phi$. After integration by parts a boundary term $\frac{\partial \mathcal{L}}{\partial (\partial_x \phi)} \delta \phi \big|_{x=0}$ remains. This term, together with the usual boundary variation, will determine the boundary condition for the field variable $\phi$.

\paragraph*{Integrable Boundary Conditions.}

In order to preserve integrability, the boundary potential has to satisfy the non-trivial constraint \eqref{eqn-BdrCondHigherCharge}, which depends on the form of the bulk Lagrangian. Hence, for a given bulk Lagrangian, the integrability constraint restricts the form of the boundary Lagrangian. In fact, following the procedure of Ghoshal and Zamolodchikov for the sine-Gordon model \cite{Ghoshal:1993tm}, we have found that in the case of the bulk $T\bar T$-Lagrangian the existence of a higher conserved charge requires the boundary potential to be zero. Our results are summarized in \tabref{tab:boundaryconds}.
\begin{table}[t]
\renewcommand{\arraystretch}{1.2}
\begin{center}
\begin{tabular}{|l||l|l|}\hline
Tested&Bulk&Boundary
\\\hline
\multirow{2}{15mm}{Generic}& \multirow{2}{30mm}{Free theory}&Mass term type: $\theta=g\phi^2/2$
\\
&&Sine--Gordon type: $\theta=\mathfrak{c}_1 \cosh(\mathfrak{c}_2 \phi-\phi_0)$
\\\hline
Order $\order{\lambda}$&$T\bar T$-deformed free theory& $\theta(\phi) \Rightarrow \theta(\phi)=0$, \quad $\theta(\phi) \Rightarrow \theta(\phi, \partial_y\phi)\neq 0$
\\\hline
Order $\order{\lambda}$&$T\bar T$-deformed Sine--Gordon& $\theta(\phi) \Rightarrow \theta(\phi)=0$
\\\hline
\end{tabular}
\end{center}
\caption{Given a certain bulk theory, we display the boundary conditions that are compatible with the existence of a higher conserved charge, which is a necessary requirement for integrability. Here in the last two rows we have performed the analysis up to the leading deformation at $\order{\lambda}$. The $\theta(\phi)$ or $\theta(\phi, \partial_y\phi)$ in the right column indicates our assumptions on the dependence of the boundary Hamiltonian $\theta$ on the fields.}
\label{tab:boundaryconds}
\end{table}

%%%%%%

\subsection{Undeformed Free Scalar and Compatible Boundary Function}

In order to illustrate how to obtain constraints on the boundary function using \eqref{eqn-BdrCondHigherCharge}, let us first study the free theory for a single massless scalar field, defined on the $-x$ axis:
\beq
S(\phi)  = \frac{1}{2} \int_{-\infty}^\infty \rd y \int_{-\infty}^0 \rd x\,  \partial_\mu \phi \partial^\mu \phi + \int dy\, \theta(\phi_\bdr) \, .
\eeq
Here $\phi_\bdr = \phi(x=0)$.
Our goal is to find the most general boundary potential $\theta$ that is compatible with \eqref{eqn-BdrCondHigherCharge}.

In order to obtain the equations of motion we consider a field variation $\phi \rightarrow \phi +\delta \phi$:
\begin{equation}
\begin{aligned}
\delta S & = \int_{-\infty}^\infty dy \int_{-\infty}^0 dx\, \partial_\mu \phi \partial^\mu \delta \phi + \int dy \frac{\delta \theta}{\delta \phi} \delta \phi \\
& = \int_{-\infty}^\infty dy \int_{-\infty}^0 dx\, (-\partial^2 \phi) \delta \phi + \int_{-\infty}^\infty dy \, \partial_x \phi \delta \phi \Big|_{x=0}+ \int dy \frac{\delta \theta}{\delta \phi} \delta \phi \, .
\end{aligned}
\end{equation}
As usual, we assume vanishing fields at infinity, such that the bulk and boundary equation of motion are given by
\begin{equation} \label{eqn-EOMfreeScalar}
\partial^2 \phi =0, \qquad \qquad \partial_x \phi +\frac{\delta \theta}{\delta \phi} \Big|_{x=0} =0 \, .
\end{equation}

\paragraph*{Conserved Higher Charges.}

For the free scalar the definition of higher charges is actually ambiguous. For instance, using the equation of motion  $\partial \bar{\partial} \phi =0$, where we remind of our complex coordinates with $\partial=\half(\partial_x-i\partial_y)$ and
$\bar\partial=\half(\partial_x+i\partial_y)$, one immediately sees that any linear combination of the form
\beq \label{eqn-TgeneralFreeScalar}
T_s  = \sum_{j=0}^s c_{s,j} (\partial^{s-j} \phi) (\partial \phi)^j
\eeq
defines an on-shell spin-$s$ conserved current in the bulk. Here the $c_{s,j}$ are constant coefficients.
However, not all of these linear combinations are conserved for any boundary Lagrangian.
We shall see how different choices of the boundary potential fix this ambiguity.

%%%%%%%%%%%%%%%%%%%%%%%%%%%%%%%%%%%%%%%%%%%%%%%%%%%

Among the conserved charges \eqref{eqn-TgeneralFreeScalar},
the $T_2$ term is the stress-energy tensor, while $T_3$ amounts to a total $z$-derivative. Therefore, the first non-trivial conserved higher charge is $T_4$.
Comparing with \eqref{eqn-TgeneralFreeScalar}, one finds that
the most general form of $T_4$ can be obtained from the integer partition of $4$ by replacing the following sets by the respective derivative terms:
\beq
 \{4\},\{3,1\},\{2,2\},\{2,1,1\},\{1,1,1,1\} \, .
\eeq
Thus, we obtain the ansatz
\beq
T_4 = c_1 \partial^4 \phi + c_2 (\partial \phi) \partial^3 \phi +  c_3 (\partial^2 \phi)^2 +  c_4 (\partial \phi)^2 \partial^2 \phi + c_5 (\partial \phi)^4 \, ,
\eeq
where the $c_j$ denote general coefficients. The conjugate $\bar{T}_4$ is obtained by the replacement $\partial \rightarrow \bar{\partial}$. Since we are dealing with free theories, the conservation of $T_4$ immediately follows from the equation of motion, $\partial \bar{\partial}\phi = 0$.

%%%%

Before going into further details, let us discuss the structure of $T_4$. The $c_1$ and $c_4$ terms are different from the others, since they are total $z$-derivatives. Therefore, it suffices to consider the case where $c_1 = c_4 =0$. (We will see that those terms drops out automatically, if we take them into account.) The $c_2, c_3$ terms are not independent, but related by an integration by parts, $(\partial^2 \phi)^2 = \partial(\partial \phi \partial^2\phi) - \partial\phi \partial^3 \phi$, so the only physical parameter is $c_3 -c_2$. We shall see that this is indeed the case.

\paragraph{Integrability Constraints on the Boundary Potential.}

For free theories,
the $\Theta_2, \bar{\Theta}_2$ terms vanish (cf.\ \eqref{eq:Thetas}). Based on this, we can compute the boundary contribution \eqref{eqn-BdrCondHigherCharge} when $r=3$,
\begin{equation} \label{eqn-T4FreeScalarBdr}
\begin{aligned}
- 8 \ii (T_4 - \bar{T}_4) & = A(\phi) (\partial_y \phi)^3 +B(\phi) \partial_y \phi \partial^2_y\phi + C(\phi) \partial^3_y \phi + (\textrm{functions of}\ \phi)\ \partial_y \phi \, ,
\end{aligned}
\end{equation}
where (for $\theta^{(j)}(\phi) = \partial^j_\phi \theta(\phi)$)
\begin{equation}
\begin{aligned}
A(\phi) & = -8 c_1 \theta^{(4)}(\phi )-4 c_4 \theta^{(3)}(\phi )-2 c_4 \theta''(\phi )-4 c_5 \theta'(\phi), \\
B(\phi) & = -24 c_1 \theta^{(3)}(\phi )-4 c_2 \theta''(\phi )-8 c_3 \theta''(\phi )-4 c_4 \theta'(\phi ),\\
C(\phi) & = -8 c_1 \theta''(\phi )-4 c_2 \theta'(\phi).
\end{aligned}
\end{equation}
Here, we have replaced $\partial_x^2$ using the bulk equation of motion, and we have replaced $\partial_x$ using the boundary equation of motion, see \eqref{eqn-EOMfreeScalar}. Employing those equations we can thus eliminate all $x$-derivatives at the boundary.

The last term of \eqref{eqn-T4FreeScalarBdr} is automatically a total $y$-derivative. For the first three terms, we can perform integration by parts on the $A,C$ terms to bring them into the form of the $B$ term:
\begin{equation}
\begin{aligned}
A(\phi) (\partial_y \phi)^3 & = \frac{d}{dy}\brk*{\int \dd \phi A(\phi)} (\partial_y \phi)^2 = - 2 \brk*{\int \dd \phi A(\phi)} \partial_y \phi \partial^2 \phi + \textrm{total $y$-derivative terms}, \\
C(\phi) \partial_y^3 \phi & = C(\phi) \frac{d}{dy} (\partial^2_y \phi) = - C'(\phi) \partial_y \phi \partial^2_y \phi + \textrm{total $y$-derivative terms}. \\
\end{aligned}
\end{equation}
Therefore, the condition of being a total $y$-derivative is equivalent to
\beq
2\int d\phi\ A(\phi) - B(\phi) + C'(\phi) = \textrm{const}.
\eeq
We allow for a constant since
\beq
\textrm{const} \times \partial_y \phi (\partial_y^2 \phi) = \textrm{const} \times\frac{1}{2} \partial_y \Big[(\partial_y \phi)^2 \Big] \,
\eeq
can be also expressed as a total derivative.
Using the explicit expressions for $A(\phi)$, $B(\phi)$, and  $C(\phi)$, we find
\begin{equation}
\begin{aligned}
2\int \dd \phi\ A(\phi) - B(\phi) +C'(\phi)= \textrm{const} & \Rightarrow (c_3-c_2) \theta''(\phi)-c_5 \theta(\phi)= \textrm{const}.
\end{aligned}
\end{equation}
We see that the dependence on $c_1$ and $c_4$ drops out automatically, and indeed the result only depends on $c_3-c_2$.
The solutions of this differential equation for the boundary function $\theta$ fall into three categories:
\begin{itemize}
\item $c_3-c_2, c_5 \neq 0$: The constant term on the right hand side just shifts the boundary potential by $\textrm{const}/c_5$, which has no physical consequences. Therefore, it suffices to take this constant to be zero, and the solution reads
\begin{equation}
    \theta(\phi) = \mathfrak{c}_1 \cosh\left( \mathfrak{c}_2 \phi - \phi_0 \right),  \quad \mathfrak{c}_2 = \sqrt{\frac{c_5}{c_3-c_2}},
\end{equation}
where $\mathfrak{c}_1, \phi_0$ denote constants of integration. This result essentially represents the boundary potential for the Sine-Gordon theory as found by Ghoshal and Zamolodchikov \cite{Ghoshal:1993tm} .
\item $c_5 =0, c_3-c_2 \neq 0$: The constant term on right hand side is now important, since it yields a quadratic contribution to the potential:
\beq
\theta(\phi)
= \frac{\textrm{const}}{2(c_3-c_2)} \phi^2 + \mathfrak{c}_3 + \mathfrak{c}_4 \phi \, ,
\eeq
where $\mathfrak{c}_3, \mathfrak{c}_4$ denote constants of integration. Physically they are not important, since we can absorb them by a constant field shift. This is the boundary function discussed in \cite{Corrigan:2004xk}.
\item $c_3 -c_2 =0, c_5 \neq 0$ case: the boundary potential is a constant. Consequently, the bulk field satisfies Neumann boundary conditions.
\end{itemize}
This completes the story of the free theory.
%
%%%%%%%%%%%%%%%%%%%%%%%%%%%%%%%%%%%%%%%%%%%%%%%%%%
%
In summary, based on the analysis of the first non-trivial higher charge $T_4$, there are only two options to choose a boundary potential (or boundary Lagrangian), which preserves integrability of the free bulk scalar:
\begin{enumerate}[1)]
  \item
   If on the one hand $\theta(\phi) = g\phi^2/2$, we find for the higher conserved charges
   \beq
  \theta(\phi) = g\frac{\phi^2}{2} \qquad \Rightarrow \qquad T_{2s} = (\partial^s \phi)^2 \, .
      \label{eq:BoundPot1}
   \eeq
   \item
   If  on the other hand $\theta(\phi) = \mathfrak{c}_1 \cosh\left( \mathfrak{c}_2 \phi - \phi_0 \right)$, the higher conserved charges take a different form:
   \begin{equation}
      \theta(\phi) = \mathfrak{c}_1 \cosh\left( \mathfrak{c}_2 \phi - \phi_0 \right) \qquad \Rightarrow \qquad T_{2s} = (\partial \phi)^{2s} +  \frac{1}{\mathfrak{c}_2^2}\ (\partial^s \phi)^2\, .
          \label{eq:BoundPot2}
   \end{equation}
\end{enumerate}
In the following it will be convenient to distinguish between charges of the form a) $(\partial^s \phi)^2$ or b) $(\partial \phi)^{2s}$. Statements about their linear combinations, as e.g.\ in \eqref{eq:BoundPot2} follow straightforwardly.
%%%%%%%

\subsection{$T\bar{T}$ Deformed Free Massless Scalar}

Now we proceed with a similar investigation in the context of deformed theories. We shall restrict to the $T\bar{T}$ case in this section.
%%%%%%%%%%%%%%%%%%%%%%%%%%%%%%%%%%%%%%%%%%%%%%%%%%

\paragraph{Bulk $T\bar{T}$ Deformation.}
As a starting point for the $T\bar T$ deformation consider the stress tensor.
For the free theory, using the general formula
\beq
T_{\mu \nu} = \frac{\partial \mathcal{L}}{\partial (\partial_\mu \phi)} \partial_\nu \phi - \delta_{\mu \nu} \mathcal{L},
\eeq
we find
\beq
\det T_{\mu \nu}^\bulk = \frac{1}{4} [(\partial_x \phi)^2 + (\partial_y \phi )^2 ]^2 \, .
\eeq
Equivalently, in complex coordinates we have
\beq
\det T_{\mu \nu}^\bulk = (\partial \phi \bar{\partial}\phi)^2 \, .
\eeq
We assume that in the bulk the action is deformed by the conventional $T\bar{T}$ deformation, \emph{i.e.}\ the deformation is defined by the equation
\beq
\frac{\rd S_\bulk}{\rd \lambda} =-\int_{-\infty}^\infty \dd y \int_{-\infty}^0 \dd x\,\det T_{\mu \nu}^\bulk\, .
\eeq
%%%%%%%%%%%%%%%
The deformed bulk Lagrangian is the Nambu-Goto Lagrangian \cite{Cavaglia:2016oda}
\beq
\mathcal{L}_\bulk = \frac{1}{2\lambda} \left(\sqrt{1+4\lambda (\partial \phi \bar{\partial}\phi)} -1 \right) = (\partial \phi \bar{\partial}\phi) - \lambda (\partial \phi)^2 (\bar{\partial} \phi)^2 + \mathcal{O}(\lambda^2) \, .
\eeq
It induces the deformed bulk equation of motion
\beq
\partial \bar{\partial} \phi = \lambda \frac{\bar{\partial}^2 \phi (\partial \phi)^2+ \partial^2 \phi (\bar{\partial} \phi)^2}{1+2\lambda \partial \phi \bar{\partial} \phi},
\eeq
and the deformed boundary equation of motion%
\footnote{We have rescaled $\theta_\lambda$ by a factor of $1/2$, to cancel the factor $1/2$ from the first term.}
\beq
0 = \frac{\partial_x \phi}{\sqrt{1+ \lambda [(\partial_x \phi)^2 + (\partial_y \phi )^2 ] }}  + \frac{\rd \theta_\lambda}{\rd \phi} \Big|_{x=0} \, .
\eeq
Here $\theta_\lambda$ represents the deformed boundary potential, and we assume it is still only a function of the bulk field, but not of its derivatives.

In the following we will identify the choices of the boundary potential which preserve integrability,
\emph{i.e.}\ which imply that all higher conserved charges satisfy the modified boundary condition \eqref{eqn-BdrCondHigherCharge}.

%%%%%%%%%%%

\paragraph*{Deformed Higher Charges.}

Using the deformed equation of motion and
the different possibilities for the initial (undeformed) higher charges $T_{2s}(\lambda=0)$, one can obtain the deformed higher charges:

\begin{enumerate}[a)]
    \item If the initial charges are $T_{2s}(0) = (\partial \phi)^{2s}$ and $\Theta_s(0) = 0$, the all-order deformations are known  and given by \cite{Cavaglia:2016oda}
    \beq \label{eqn-DeformedFreeCharge}
    T_s(\lambda) = \frac{(\partial \phi)^s}{\mathcal{S}} \left( \frac{2}{\mathcal{S}+1}  \right)^{s-2}\, , \quad
    \Theta_{s-2}(\lambda) = \frac{\lambda (\partial \phi)^s (\bar{\partial} \phi)^2}{\mathcal{S}} \left( \frac{2}{\mathcal{S}+1}  \right)^{s}\, ,
    \eeq
    where $\mathcal{S} = \sqrt{1+4\lambda \partial \phi \bar{\partial}\phi}$.
    \item On the other hand, if the initial charges are
     $T_{2s}(0) = (\partial^s \phi)^2$ and $\Theta_s(0) = 0$, the all order deformations can in principle be obtained, but there is no simple formula for a generic deformed charge. The leading higher charges $T_4(\lambda)$ and $\Theta_2(\lambda)$ are given by \cite{Conti:2019dxg}
    \begin{equation} \label{eqn-DeformedT4Type2}
        \begin{aligned}
            T_4(\lambda) & = \frac{(\partial \phi)^2}{\mathcal{S}} \brk*{\frac{(\mathcal{S}-1)^4 \bar{\partial}^2 \phi -16 \lambda^2 (\bar{\partial}\phi)^4 \partial^2 \phi}{4\lambda (\mathcal{S}-1)(\mathcal{S}^2+1)(\bar{\partial}\phi)^3} }^2, \\
            \Theta_2(\lambda) & = \frac{(\mathcal{S}-1)^2}{4\lambda\mathcal{S}} \brk*{\frac{(\mathcal{S}-1)^4 \bar{\partial}^2 \phi -16 \lambda^2 (\bar{\partial}\phi)^4 \partial^2 \phi}{4\lambda (\mathcal{S}-1)(\mathcal{S}^2+1)(\bar{\partial}\phi)^3} }^2.
        \end{aligned}
    \end{equation}
\end{enumerate}
The conjugates $\bar{T}_s$ and $\bar{\Theta}_s$ are obtained by replacing $z$ by $\bar{z}$.

Clearly, if the initial charges are linear combinations of both types a) and b) presented above as in the case \eqref{eq:BoundPot2}, the deformed charges are given by the same (deformed) linear combinations.

%%%%%%%%%%%%%%%%
\paragraph{Boundary $T\bar{T}$ Deformation for $\theta=\theta(\phi)$.}

Let us now investigate the following boundary integrablity condition \eqref{eqn-BdrCondHigherCharge} for the first few charges:
\begin{equation}
-\ii(T_{r+1}-\bar T_{r+1} +\bar \Theta_{r-1} - \Theta_{r-1})\Big|_{x=0} = \frac{\rd}{\rd y} \theta_r(y).
\end{equation}

%%%%%

Even though conservation of $T_2$ does not correspond to integrability, we first discuss this case for completeness.
For $r=1$ the above equation involves the deformed stress tensor $T_2$. Note that $T_2$ is unique, such that we can use \eqref{eqn-DeformedFreeCharge}. We immediately see that $\Theta_0(\lambda)$ and $\bar{\Theta}_0(\lambda)$ cancel out, and the non-trivial contribution at the boundary is given by
\beq
\begin{aligned}
-\ii (T_{2} + \bar{\Theta}_{0}-\bar{T}_2 - \Theta_{0})\Big|_{x=0} & = \frac{\ii}{\sqrt{1+4\lambda \partial \phi \bar{\partial}\phi }} \Big( (\partial \phi)^2 - (\bar{\partial} \phi)^2\Big) \\
& = \frac{\partial_x \phi \partial_y \phi}{\sqrt{1+4\lambda \partial \phi \bar{\partial}\phi }} \Big|_{x=0}\\
& = -\partial_y \phi\ \frac{\rd \theta_\lambda}{\rd \phi} \, .
\end{aligned}
\eeq
It is now clear that as long as $\theta_\lambda$ is a function of $\phi$ only, the expression above is a total $y$-derivative, regardless of the functional form of $\theta_\lambda$.
%%%%%

The first non-trivial constraint on the boundary function $\theta_\lambda$ comes from the leading higher charge $T_4$.
A generic all-order analysis seems rather involved, so we restrict to a perturbative analysis
at $\mathcal{O}(\lambda)$. The bulk equation of motion is then given by
\beq
\partial_x^2 \phi = - \partial^2_y \phi + \lambda \Big( \partial^2_y \phi \big[(\partial_y \phi)^2 - (\partial_x \phi)^2 \big]  + 2 \partial_x \phi \partial_y \phi (\partial_{x} \partial_y \phi) \Big) + \mathcal{O}(\lambda^2) \, .
\eeq
Similarly, we can solve the deformed boundary equation of motion. Expanding the boundary potential in $\lambda$ according to
\beq
\theta_\lambda = \theta_{(0)} + \lambda \theta_{(1)} + \mathcal{O}(\lambda^2),
\eeq
we find
\beq
\partial_x \phi \big|_{x=0} = - \theta_{(0)}' - \lambda \brk*{\theta_{(1)}'- \sfrac{1}{2} \theta_{(0)}' [(\theta_{(0)}')^2 + (\partial_y \phi)^2] } + \mathcal{O}(\lambda^2).
\eeq
Using those deformed equations of motion, we can analyze the defomed $T_4$ for different undeformed charges:
\begin{enumerate}[a)]
    \item  If the undeformed charges are $T_4 = (\partial \phi)^4$, using the general expressions \eqref{eqn-DeformedFreeCharge}, we find that at $\mathcal{O}(\lambda)$ the expression only contains single $x$-derivatives.
    Substituting $\partial_x \phi$,  we obtain the $\mathcal{O}(\lambda)$ contributions
\beq
\begin{aligned}
& -2 \ii (T_{4} + \bar{\Theta}_{2}-\bar{T}_4 - \Theta_{2} ) \Big|_{x=0, \mathcal{O}(\lambda)} \\
& = \Big(\frac{3}{4}(\theta_{(0)}')^3 - \theta_{(1)}' \Big) (\partial_y \phi)^3 + \frac{3}{8} \theta_{(0)}' (\partial_y \phi)^5 + (\textrm{functions of}\ \phi)\ \partial_y \phi.
\end{aligned}
\eeq
\item On the other hand, if the undeformed charge reads $T_4= (\partial^2 \phi)^2$, using \eqref{eqn-DeformedT4Type2} and
substituting $\partial_x \phi, \partial_x^2 \phi$, we find that the additional $\mathcal{O}(\lambda)$ contribution is given by
\beq \label{eqn-T4Type2OneLoop}
\begin{aligned}
& -2 \ii (T_{4} + \bar{\Theta}_{2}-\bar{T}_4 - \Theta_{2} ) \Big|_{x=0, \mathcal{O}(\lambda)} \\
& = (\textrm{functions of}\ \phi)\ \partial_y \phi + \Big(\frac{3}{2} \theta_{(0)}' (\theta_{(0)}'')^2 (\partial_y \phi)^3 - \big[ \frac{3}{2} (\theta_{(0)}')^2 \theta_{(0)}'' + 2 \theta_{(1)}''\big] \partial_y \phi \partial^2_y \phi\Big) \\
& \qquad + \frac{3}{2} \Big( \theta_{(0)}'' (\partial_y \phi)^3 \partial^2_y \phi - \theta_{(0)}'(\partial_y \phi) (\partial^2_y\phi)^2 \Big).
\end{aligned}
\eeq
\end{enumerate}
From the expressions for both of the above cases one immediately sees that the five-derivative terms, namely the terms that are proportional to $(\partial_y\phi)^5,(\partial_y \phi) (\partial^2_y\phi)^2$, or $(\partial_y \phi)^3 (\partial^2_y\phi)$ only depend on the undeformed boundary potential $\theta_{(0)}$. Since we are missing a term $\partial_y^3 \phi \partial^2_y \phi$ to combine with $\partial_y \phi (\partial^2_y \phi)^2$, we are unable to turn those five-derivative terms into a total $y$-derivative.
The only solution which preserves integrability is to set $\theta_{(0)}$ to zero.

If we set  $\theta_{(0)}=0$, the $\theta_{(1)}$ dependent terms now have exactly the same structure as those for the boundary function in the undeformed theory. Hence, the most general solution at this order
is identical to the undeformed boundary function. In other words, the boundary potential would be delayed by one order. However, since we have found that the $\order{\lambda}$ analysis forces $\theta_{(0)}$ to vanish, it is conceivable that going to the next perturbative order $\order{\lambda^2}$ implies that also $\theta_{(1)}$ has to vanish in order to preserve integrability.
This phenomenon is actually quite general. We have verified that for the Sine-Gordon theory, the deformed boundary potential also gets delayed, see appendix \ref{apd-SG} for details. This suggests that the most general boundary potential which is compatible with integrability and the bulk $T\bar{T}$ deformation is zero.

%%%%%%

%%%%%%%%%%%%%%%%
\paragraph{Boundary $T\bar{T}$ Deformation for $\theta=\theta(\phi, \partial_y \phi)$.}

The result in the previous paragraph does not necessarily exclude the possibility that the boundary potential depends on the derivatives of $\phi$. In fact, we shall see that at leading order, the deformed boundary Lagrangian can be non-trivial, if it depends also on $\partial_y \phi$.

Drawing inspiration from the $T\bar{T}$ deformed charges, we now assume that the deformed boundary potential is a function of $\phi$ and $\partial_y \phi$. The deformed Lagrangian takes the form
\beq
\theta_\lambda = \theta_{(0)}(\phi) + \lambda \Big(\theta_{(1,1)}(\phi) + \theta_{(1,2)}(\phi) (\partial_y \phi)^2 \Big) + \mathcal{O}(\lambda^2),
\eeq
where we assume the undeformed boundary Lagrangian $\theta_{(0)}$ is still of the potential type, \emph{i.e.}\ does not depend on derivatives of $\phi$. Here, we have assumed that the $\mathcal{O}(\lambda)$ terms only contain $y$-derivatives up to second order $\partial_y^2 \phi$, because higher order derivative terms would contribute to \eqref{eqn-BdrCondHigherCharge} with more than five-derivative terms, which would be inconsistent with the structure we have found.
Now, the deformed boundary equation of motion becomes
\begin{align}
\partial_x \phi \Big|_{x=0} = - \theta_{(0)}' - \lambda \Big(
&\theta_{(1,1)}(\phi)' - \theta_{(1,2)}(\phi)' (\partial_y \phi)^2-2\theta_{(1,2)}(\phi)(\partial_y \phi)^2
\nonumber\\
&\qquad+\sfrac{1}{2} \theta_{(0)}' [(\theta_{(0)}')^2 + (\partial_y \phi)^2] \Big) + \mathcal{O}(\lambda^2).
\end{align}
With the additional $\theta_{(1,2)}$ term, there will be a new five-derivative contribution proportional to $(\partial_y^3 \phi) (\partial^2_y \phi)$ in \eqref{eqn-T4Type2OneLoop}. Combined with the problematic $\partial_y \phi (\partial^2_y \phi)^2$ term, they can become a total $y$-derivative.
Explicitly, they read
\beq
(\partial_y \phi) (\partial^2_y \phi)^2 \underbrace{(8 \theta_{(1,2)}'-\frac{3}{2} \theta_{(0)}')}_{= \mathcal{A}_{1,2,2}} +(\partial_y^3 \phi) (\partial^2_y \phi) \underbrace{4 \theta_{(1,2)}}_{= \mathcal{A}_{2,3}}.
\eeq
Thus, the coefficients must satisfy
\beq
\mathcal{A}_{2,3}(\phi) - 2\int \dd \phi\ \mathcal{A}_{1,2,2}(\phi) = \text{const}\Rightarrow \theta_{(1,2)} = \mathfrak{c}_{1,2} + \frac{1}{4} \theta_{(0)}\, ,
\eeq
where $\mathfrak{c}_{1,2}$ denotes an arbitrary constant. What remains in \eqref{eqn-T4Type2OneLoop} is simply a term of the form $2 \theta_{(0)}'' (\partial_y \phi)^3 (\partial^2_y \phi)$.
Now, we can use the explicit forms of the undeformed boundary potential, which are compatible with integrability (see \eqref{eq:BoundPot1} and \eqref{eq:BoundPot2}):
\begin{enumerate}[1)]
    \item If the undeformed boundary potential reads $g\phi^2/2$, the remaining term $2 \theta_{(0)}'' (\partial_y \phi)^3 (\partial^2_y \phi)$ is automatically a total $y$-derivative. Demanding the three-derivative terms to be a total $y$-derivative, we find
    \beq
    \theta_{(1,1)} = - \frac{5g^3}{32} \phi^4 + \mathfrak{d}_1 \phi^2 + \mathfrak{d}_2 \phi +\mathfrak{d}_3,
    \eeq
    where the $\mathfrak{d}_j$ denote constants.
    \item If the undeformed boundary potential reads $\mathfrak{c}_1 \cosh(\mathfrak{c}_2 \phi - \phi_0)$, the term $2 \theta_{(0)}'' (\partial_y \phi)^3 (\partial^2_y \phi)$ must combine with other five-derivative terms coming from the deformed $(\partial \phi)^4$ charge. For the five-derivative terms the condition of being a total derivative is then automatically satisfied.
    Plugging this into the third-order terms, we find a second-order differential equation for $\theta_{(1,1)}$:
    \beq
    0 = \frac{3}{2} \mathfrak{c}_1^2 \mathfrak{c}_2^2 \sinh ^2\left(\mathfrak{c}_2 \phi - \phi_0 \right) \left(\mathfrak{c}_{1,2}-2 \mathfrak{c}_1 \cosh \left(\mathfrak{c}_2 \phi - \phi_0 \right)\right)-\frac{2 \theta _{1,1}''(\phi )}{\mathfrak{c}_2^2}.
    \eeq
    The solution to this equation is easily obtained, but it is not illuminating. Therefore we shall not present it here.
\end{enumerate}

\paragraph*{Summary.}

To summarize, given the $T\bar T$ deformed bulk Lagrangian our $\mathcal{O}(\lambda)$ perturbative analysis shows the following:
\begin{itemize}
    \item If the boundary function only depends on $\phi$ and not on its derivatives, the deformed boundary potential is delayed for one order. This suggests that for the full $T\bar{T}$ deformed bulk Lagrangian, the only integrability-preserving boundary potential is zero.
    \item If instead we allow for derivative corrections at $\mathcal{O}(\lambda)$ in the boundary function, then a non-trivial, integrability-preserving solution for the deformed boundary Lagrangian exists. This suggests that a non-trivial, integrability-preserving boundary Lagrangian necessarily depends on $\partial_y \phi$.
\end{itemize}

%%%%%%%%%%%%%%%%%%%%%%%%%%%%%%%%%%%%%%%%%%
\section{Review of Deformations for Closed \& Open Boundaries}
\label{sec:GeneralDefs}

In this section we discuss some general principles underlying deformations of two-dimensional models, \emph{e.g.}\ field theories, spin chains or the Bose gas.

%%%
\subsection{Closed Boundaries}
\label{sec:DefsClosedBounds}
Let us review the construction of \cite{Bargheer:2008jt,Bargheer:2009xy} adapted to the field theory context.
We are interested in deformations of a Hamiltonian, or more generically, a set of conserved charges, which preserve
locality%
\footnote{This should not be confused with locality of quantum field theory in the Wightman axioms \cite{Streater:1989vi}.}
in the sense that the undeformed as well as the deformed Hamiltonian can be written as an integral over a local density $\hamdens(x)$:
\begin{equation}
\ham=\int \dd x \,\hamdens(x).
\label{eq:Hamloc}
\end{equation}
In the following we will focus on continuous models but similar considerations apply to discrete spin chains, where integrals are replaced by lattice sums.
A general class of deformed Hamiltonians $\ham_{\lambda}$ is defined via a parallel transport equation of the following form:
\begin{equation}
\frac{\dd}{\dd \lambda} \ham_{\lambda}=\comm{\genX}{\ham_{\lambda}}.
\label{eq:deformationequationH}
\end{equation}
where the deformation operator $X$ also depends on $\lambda$, but we omit it systematically for simplicity. For an integrable model, such deformations preserve integrability if the integrable charges $\charge_r$ are deformed by means of the same deformation equation as the Hamiltonian.
Deformations that preserve locality in the above sense are conveniently introduced using the notion of \emph{bilocal operators} defined as
\begin{align}
\biloc{A}{B}_\text{closed}
&:=\int_{s_\Left}^{s_\Right}\dd x_2 \int_{s_\Left}^{x_2} \dd x_1  \, \mathcal{A}(x_1)\mathcal{B}(x_2)
=\includegraphicsbox[scale=.42]{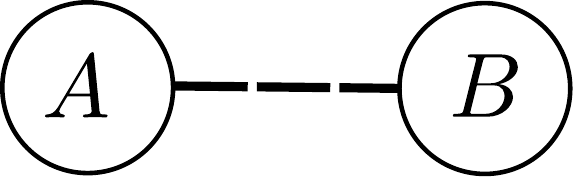}\,\,.
\label{eq:bilocfieldclosed}
\end{align}
Here $A$ and $B$ represent two local operators in the above sense
and $s_\Left$ and $s_\Right$ denote the positions of the left and right boundary, respectively.  For the infinite line we have $s_{\LeftRight}=\mp \infty$.
The label `closed' indicates that we will slightly refine this definition of a bilocal operator in the context of boundary systems.
While the introduction of the above bilocal operators may seem ad hoc at first sight, it is motivated by the fact that it has been shown that for spin chains this class of deformation generators exhausts the complete space of integrability preserving deformations found for closed \cite{Beisert:2005wv} and open \cite{Beisert:2008cf} $\alg{gl}(N)$ chains, as well as in the XXZ case \cite{Beisert:2013voa}.

\begin{figure}
\begin{center}
$\comm{\biloc{\charge_r}{\charge_s}}{\charge_t}\quad=\quad$
\includegraphicsbox[scale=.4]{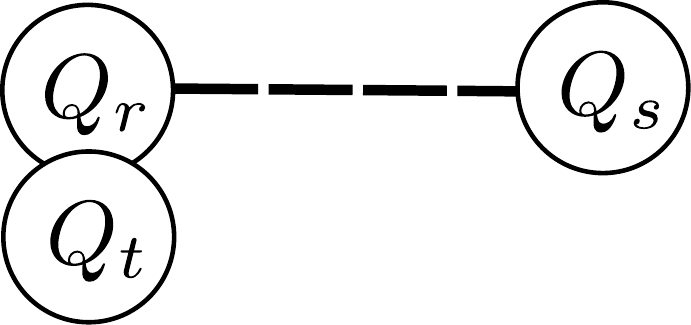}
\quad$+$\quad
\includegraphicsbox[scale=.4]{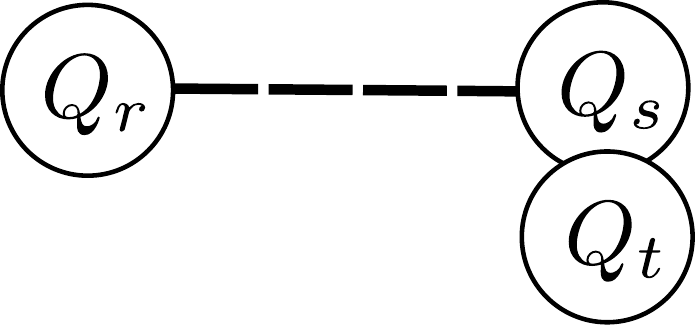}
\quad$+$\quad
\includegraphicsbox[scale=.4]{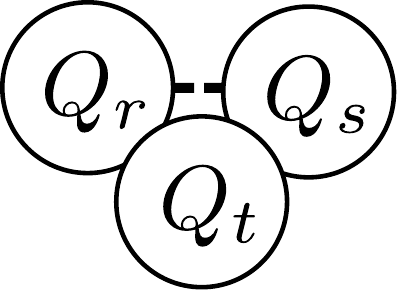}
\caption{The only non-vanishing contribution to the commutator of a bilocal charge with a local charge originates from the last term where both legs $\charge_r$ and $\charge_s$ of the bilocal operator are close to each other. The first two terms vanish since the local charge $\charge_t$ commutes with both local charges $\charge_r$ and $\charge_s$.}
\end{center}
\label{fig:biloccomm}
\end{figure}

Importantly, locality of the Hamiltonian $\ham$ defined via \eqref{eq:deformationequationH} is preserved if the local operators $A$ and $B$ both commute with $\ham$, e.g.\ for two conserved charges $A=\charge_r$ and $B=\charge_s$. In that case, \emph{i.e.}\ for $\genX=\biloc{\charge_r}{\charge_s}$, the only non-vanishing contributions to the commutator $\comm{\biloc{A}{B}}{\ham}$ originate from the local term
$\chargedens_r(x)\chargedens_s(x)$ in the definition of the bilocal operator \eqref{eq:bilocfieldclosed}, which yields a local result, see \figref{fig:biloccomm}.%
\footnote{Note that e.g.\ for a \emph{tri}-local operator this would not be the case.}
Would either $\charge_r$ or $\charge_s$ not commute with $\ham$, the result of the commutator were bilocal.
This subtlety arises in the case of open boundaries where the parity-odd charges are typically not conserved but still required to generate the full space of admissible deformations \cite{Loebbert:2012yd}, see \secref{sec:openboundaries}. Note that the charges $\charge_r$ can be taken to be the spacetime $\mom$- or $\ham$-type charges described in \secref{sec:ConservedCharges} or some internal commuting charge, see \cite{Beisert:2013voa} for an example of latter. While the $T\bar T$-deformation belongs to the class of deformations induced by bilocal spacetime charges, the combination of spacetime and internal charge is referred to as $J\bar T$-deformations in the field theory context,
see \cite{Guica:2017lia,Guica:2019vnb,Anous:2019osb,Nakayama:2018ujt,Frolov:2019xzi,Aguilera-Damia:2019tpe,Chakraborty:2019mdf,Apolo:2018qpq}. Similar deformation for spin chains were considered in \cite{Beisert:2013voa}.

Note that also the constituent charges $\charge_r$ and $\charge_s$ of the bilocal charge $\biloc{\charge_r}{\charge_s}$ should be deformed via an equation of the form \eqref{eq:deformationequationH}, in order to preserve $\comm{\ham(\lambda)}{\charge_r(\lambda)}=0$ for $\lambda\neq0$. That is, strictly speaking all operators in this paragraph carry an argument $\lambda$, which we have omitted to avoid clutter.

%%%%%%
\paragraph{Relation to $T\bar T$-like Deformations.}

Let us relate the above bilocal deformations to bilinear deformations expressed in terms of currents, see \cite{Pozsgay:2019xak,Pozsgay:2019ekd,Marchetto:2019yyt}.
Consider the deformation equation \eqref{eq:deformationequationH} with
\begin{align}
\genX=\biloc{\charge_r}{\charge_s}=\int_{x_1<x_2}\dd x_1 \dd x_2\,\chargedens_r(x_1)\chargedens_s(x_2).
\end{align}
Using the Heisenberg equation \eqref{eq:HeisEq} given by $\partial_y\chargedens_r=\comm{\hamdens}{\chargedens_r}$ as well as the conservation  equation \eqref{eq:ConsEq} given by  $\partial_y\chargedens_r=-\partial_x \currentdens_r$, we find
\begin{align}
\comm{X}{\ham}
&=
\int_{x_1<x_2}\dd x_1 \dd x_2\brk2{
\comm{\chargedens_r(x_1)}{\ham}\chargedens_s(x_2)
+
\chargedens_r(x_1)\comm{\chargedens_s(x_2)}{\ham}
}
\nonumber\\
&=
\int_{s_\Left}^{s_\Right} \dd x_2 \brk*{\currentdens_r(x_2)-\currentdens_r(s_\Left)}\chargedens_s(x_2)
+
\int_{s_\Left}^{s_\Right} \dd x_1 \chargedens_r(x_1)
\brk*{\currentdens_s(s_\Right)-\currentdens_s(x_1)}.
\end{align}
We  introduce the operator
\begin{equation}
\Odens_{rs}=-\epsilon_{\mu\nu}\current_r^\mu \current_s^\nu=\currentdens_r \chargedens_s-\chargedens_r \currentdens_s,
\end{equation}
such that we can write
\begin{equation}
\comm{X}{\ham}
=
\int_{s_\Left}^{s_\Right}   \Odens_{rs}(x)\dd x-\currentdens_r(s_\Left)\charge_s+\charge_r \currentdens_s(s_\Right).
\end{equation}
On the infinite line with $s_\Left=-\infty$ and $s_\Right=+\infty$, where the last two terms drop out, we thus find
\begin{align}
\label{eq:defHJJ1}
\frac{\dd H_{\lambda}}{\dd\lambda}=\int_{-\infty}^{\infty} \mathcal{O}_{rs}(x)\dd x.
\end{align}
In the special case of $s=1,r=2$ with $T=T_2$, $\Theta=\Theta_0$, we have
\begin{equation}
\Odens_{21}=4\brk*{T\bar T-\Theta\bar \Theta} = \det T_{\mu \nu}.
\end{equation}
Note that the densities $\chargedens_r$ and $\currentdens_s$ also receive deformations in $\lambda$,
and that the deformation equation \eqref{eq:deformationequationH} is formally solved by
\begin{align}
H_{\lambda}=U_{\lambda}^{-1}H_0 U_{\lambda},
\qquad\qquad
U_{\lambda}=\mathcal{P}\exp\left[-\int_0^{\lambda}\genX(\lambda')\, \dd\lambda' \right].
\end{align}
%%%
\subsection{Open Boundaries}
\label{sec:openboundaries}

For systems with open boundary conditions there are important differences in the construction of locality and integrability preserving deformations to the closed case discussed above, see \cite{Loebbert:2012yd}.
Firstly, for open systems merely the parity-even charges $\charge_{2r}=\ham_{2r-1}$ are conserved, while conservation of the odd charges $\charge_{2r-1}=\mom_{2r-1}$ is typically broken by boundary terms. In particular, the odd momentum operator $\charge_1=\mom$ is not conserved. Nevertheless, we can still define odd ``charge operators", see \eqref{eq:EvenOddCharges}, by the requirement that these are conserved in the bulk of the theory. In particular, we have
\begin{equation}
\comm{\charge_{r}}{\charge_{2s+1}}=\text{boundary terms},
\label{eq:evenwithodd}
\end{equation}
where \emph{boundary terms}  may act nontrivially on the boundary  but vanish in the bulk.
Note that formally we may include boundary terms like the boundary Hamiltonian $\theta$ into the bulk density of the operators by writing them as total derivative terms, \emph{e.g.}:
\begin{equation}
\theta(x=s_\Right,y)=\int_{-\infty}^{s_\Right} \dd x \, \partial_x\theta(x,y).
\end{equation}
Note, however, that the classical analysis of the $T\bar T$-deformed model in the previous \secref{sec:ClassDef} suggests to set the boundary functions $\theta$ of the considered charges to zero.

Secondly, deformations with bilocal charges $\biloc{\charge_r}{\charge_s}$ will generically result in \emph{bilocal} deformations of the even conserved charges $\charge_{2t}$ of the open model, if one of the charges $\charge_r$ or $\charge_s$ is odd. Therefore the order of the local operators entering the bilocal operator is crucial when applied to semi-infinite systems.
We will distinguish such systems with either an open boundary on the left (left-open) or on the right (right-open).
For a left-open model for instance, we obtain \emph{local} deformations only when using the bilocal operator $\biloc{\charge_{2r}}{\charge_{2s+1}}$, but not for $\biloc{\charge_{2s+1}}{\charge_{2r}}$.

Finally, in the case of open boundaries nontrivial deformations can be induced by \emph{local operators} in addition to the above bilocal charges. In particular, this means that the precise choice of local regularization of bilocal operators becomes important.
In the open case we define the bilocal operators as%
\footnote{Note that one can also use this regularized version of bilocal operators in the case of closed boundary conditions, cf.\ \eqref{eq:bilocfieldclosed}. The additional local term makes no difference in that case.}
\begin{align}
\biloc{A}{B}
&:=\int_{s_\Left}^{s_\Right}\dd x_2 \int_{s_\Left}^{x_2} \dd x_1  \, \half \acomm{\mathcal{A}(x_1)}{\mathcal{B}(x_2)}-\quarter \int_{s_\Left}^{s_\Right} \dd x\,\acomm{\mathcal{A}(x)}{\mathcal{B}(x)}
\nonumber\\
&=\int_{s_\Left}^{s_\Right} \dd x_2\int_{s_\Left}^{x_2} \dd x_1  \, \half \brk*{1-\half \delta(x_1-x_2)}\acomm{\mathcal{A}(x_1)}{\mathcal{B}(x_2)}.
\label{eq:bilocfield}
\end{align}
Here $\acomm{\cdot}{\cdot}$ denotes the anti-commutator. The above definition yields
\begin{equation}
\biloc{A}{B}+\biloc{B}{A}=\half \acomm{A}{B}.
\end{equation}
In particular, this regularization implies that the sum of conserved bilocal charges
\begin{equation}
\biloc{\charge_{r}}{\charge_{s}}+\biloc{\charge_{s}}{\charge_{r}}=\charge_{r}\charge_{s}
\end{equation}
commutes with the Hamiltonian (and higher integrable charges) in the bulk, \emph{i.e.}\ the only non-trivial bulk deformations are induced by the difference of bilocal charges $X=\biloc{\charge_{r}}{\charge_{s}}-\biloc{\charge_{s}}{\charge_{r}}$ when inserted into \eqref{eq:bilocfield}.
Moreover, the bulk projection of the commutator of the form $\comm{\biloc{\charge_{2r}}{\charge_{2s+1}}}{\charge_{2t}}$ equals the bulk projection of the commutator $-\comm{\biloc{\charge_{2s+1}}{\charge_{2r}}}{\charge_{2t}}$, \emph{i.e.}\
\begin{equation}
\comm{\biloc{\charge_{2r}}{\charge_{2s+1}}}{\charge_{2t}}+\comm{\biloc{\charge_{2s+1}}{\charge_{2r}}}{\charge_{2t}}\Big|_\text{bulk}
=\half \comm{\acomm{\charge_{2r}}{\charge_{2s+1}}}{\charge_{2t}}\Big|_\text{bulk}
=0,
\label{eq:bilocbulkequal}
\end{equation}
which will be important for the below construction.

%%%%%%
\subsection{Explicit Construction for Open Boundaries}
Following \cite{Loebbert:2012yd}, in this section, we present more details of the deformations for open boundaries which are the main focus of this paper. In particular, we will review how deformations generated in a left- and right-open model can be combined into deformed charge operators for systems with two boundaries.

First of all we refine the above notion of boundary terms by introducing \emph{left boundary terms} $\loc_\Left^\bdr$ and \emph{right boundary terms} $\loc_\Right^\bdr$, which only act on the left or the right boundary, respectively. Acting for instance with a right boundary term  $\loc_\Right^\bdr$ on a state in a left-open model yields zero:
\begin{equation}
 \loc_\Right^\bdr \ket{\psi}_\Left=0.
\end{equation}
Here we denote states in the left- or right-open model by $\ket{\cdot}_\Left$ or $\ket{\cdot}_\Right$, respectively.
Moreover, it will be useful to introduce a notion of setting boundary terms to zero. We employ the notation $|_\Left$ to indicate that we set right boundary terms to zero and $|_\Right$ to set left boundary terms to zero.
More explicitly, if we apply the boundary conditions of a left- or right-open model denoted by $|_\Left$ or $|_\Right$, respectively, we have
\begin{align}
\loc_\Left^\bdr|_\Left&=\loc_\Left^\bdr,
&
\loc_\Left^\bdr|_\Right&=0,
\\
\loc_\Right^\bdr|_\Left&=0,
&
\loc_\Right^\bdr|_\Right&=\loc_\Right^\bdr.
\end{align}
The bulk part of a local operator $\loc$ can thus be defined as
\begin{equation}
\loc^\bulk=\loc|_\Left|_\Right
\equiv \loc|_{\Left\Right}.
\end{equation}
Note that for a system with open boundaries, the odd charges commute up to boundary terms according to
\begin{equation}
\comm{\charge_{r}}{\charge_{2s+1}}=\loc_\Left^\bdr+\loc_\Right^\bdr.
\end{equation}
%%%%%%

\paragraph{Semi-Infinite Systems.}
We now want to deform a set of even charge operators $\charge_{2r,\LeftRight}=\charge_{2r,\LeftRight}(\lambda=0)$, which are conserved in the left or right open model, respectively. For a non-integrable model this set may only contain the Hamiltonian $\charge_{2,\LeftRight}=\ham_\LeftRight$.
Accordingly, we introduce two sets of charges deformed in the parameter $\lambda$ labelled by $\Left$ and $\Right$ and defined by the equation%
\footnote{Here we refrain from adding a label $X$ to the deformation parameter $\lambda$, which can be useful when studying different types of deformations at the same time.}
\begin{equation}
\frac{\dd}{\dd \lambda} \charge_{2r,\Left/\Right}(\lambda)=\comm{X_{\Left/\Right}(\lambda)}{Q_{2r,\Left/\Right}(\lambda)}\big|_{\Left/\Right}.
\label{eq:defeqopenfieldLR}
\end{equation}
This equation guarantees that charges, which commute for $\lambda=0$, will also commute for a non-vanishing deformation parameter $\lambda\neq 0$.
Here the requirement of locality for the deformed charges implies that the bilocal deformation generators have to be chosen as
\begin{align}
X_\Left
&=
+\biloc{\charge_{2r}}{\charge_{2s+1}},
&
X_\Right
&=
-\biloc{\charge_{2s+1}}{\charge_{2r}}.
\end{align}
The two sets of (deformed) charges defined by \eqref{eq:defeqopenfieldLR} take the form
\begin{equation}
\charge_{2r,\LeftRight}=\charge_{2r}^\bulk+\charge_{2r,\LeftRight}^\bdr,
\end{equation}
with a bulk term and a term left or right boundary term, respectively.
Here the building blocks are defined in terms of the solutions of \eqref{eq:defeqopenfieldLR} according to
\begin{align}
\charge_{2r}^\text{bulk}
&=\charge_{2r,\Left}|_{\Left\Right}=\charge_{2r,\Right}|_{\Left\Right},
\\
\charge_{2r,\Left}^\bdr
&=\charge_{2r,\Left}-\charge_{2r}^\text{bulk},
\label{eq:QdefLbdr}
\\
\charge_{2r,\Right}^\bdr
&=\charge_{2r,\Right}-\charge_{2r}^\text{bulk},
\label{eq:QdefRbdr}
\end{align}
cf.\ \eqref{eq:bilocbulkequal} for the second equality in the first line.
The deformed charges
commute by definition in the left- or right-open model, respectively, e.g.\ in the left-open case we have
\begin{equation}
\comm{\charge_{2r,\Left}}{\charge_{2s,\Left}}=\loc^\bdr_\Right|_\Left=0.
\label{eq:commex}
\end{equation}
Here $\loc^\text{bdr}_\Right$ denotes some boundary term that only acts on a right boundary. In the left-open model, however, the right boundary is absent and thus the boundary term vanishes when imposing the respective boundary conditions as denoted by $|_\Left$.
The above equation \eqref{eq:commex} can be expanded according to
\begin{align}
&\comm{\charge_{2r}^\bulk}{\charge_{2s}^\bulk}
=\loc_\Left^\bdr+\loc_\Right^\bdr,
\\
&\comm{\charge_{2r}^\bulk}{\charge_{2s,\Left}^\bdr}
+
\comm{\charge_{2r,\Left}^\bdr}{\charge_{2s}^\bulk}
+
\comm{\charge_{2r,\Left}^\bdr}{\charge_{2s,\Left}^\bdr}=-\loc_\Left^\bdr.
\end{align}
%%%%%%%%%

\paragraph{Finite Systems.}
The next important step is to proceed to a finite open system with boundaries on the left \emph{and} on the right. Using the above building blocks from both half-open systems, we define deformed charges as
\begin{equation}
\charge_{2r}(\lambda)=\charge_{2r}^\text{bulk}(\lambda)+\charge_{2r,\Left}^\bdr(\lambda)+\charge_{2r,\Right}^\bdr(\lambda).
\label{eq:finiteopendefcharges}
\end{equation}
The charges defined in this way obey
\begin{equation}
\comm{\charge_{2r}}{\charge_{2s}}
=
\comm{\charge_{2r,\Left}^\bdr}{\charge_{2s,\Right}^\bdr}
+
\comm{\charge_{2r,\Right}^\bdr}{\charge_{2s,\Left}^\bdr}
=
\loc_{\Left\&\Right}^\bdr.
\end{equation}
Here the terms $\loc_{\Left\&\Right}^\bdr$ act on both boundaries at the same time and are referred to as \emph{spanning terms}, cf.\ \cite{Beisert:2008cf,Loebbert:2012yd}. In the spin chain case it is clear that the interaction range of the charge deformations increases with increasing order in $\lambda$. Hence, for a given chain of finite length, spanning terms arise at a finite order of the deformation parameter $\lambda$.
In the field theory case the deformations of the charge operators at a given perturbative order in $\lambda$ are naively localized at some point $x$ and would thus never act on both boundaries at the same time, \emph{i.e.}\ contributions $\loc_{\Left\&\Right}^\bdr$ would be zero in the field theory. In a non-perturbative context, however, interactions seeing both boundaries may arise.

Finally we note that deformations with bilocal operators $\biloc{\charge_{2r}}{\charge_{2s}}$ composed of two even charges can as well be performed within the finite model and thus correspond to trivial similarity transformations. Deformations with $\biloc{\charge_{2r+1}}{\charge_{2s+1}}$ do not result in local deformations.

%%%%%%%%%%%
\paragraph{Expressions in Terms of Currents.}
A feature that did so far not appear in the context of field theory $T\bar T$-like deformations for closed systems are the above deformations generated by odd charges, see \cite{Loebbert:2012yd} for the spin chain case.
Let us thus translate these deformations into expressions in terms of currents. For the open model the Hamiltonian does not commute with the odd charges and we have
\begin{align}
0\neq \comm{\ham}{\charge_{2r+1}}
&=\int_{s_\Left}^{s_\Right}\comm{H}{\chargedens_{2r+1}(x)}\dd x
=\int_{s_\Left}^{s_\Right}\partial_y\chargedens_{2r+1}(x)
\nonumber\\
&=\int_{s_\Left}^{s_\Right} (-\partial_x \currentdens_{2r+1}(x))
=-\currentdens_{2r+1}(s_\Right)+\currentdens_{2r+1}(s_\Left).
\end{align}
Hence, we find
\begin{align}
\comm{+\charge_{2r+1}}{\ham}|_\Left
&=
-\currentdens_{2r+1}(s_\Left),
\\
\comm{-\charge_{2r+1}}{\ham}|_\Right
&=
-\currentdens_{2r+1}(s_\Right),
\end{align}
such that via \eqref{eq:defeqopenfieldLR} the deformed Hamiltonian becomes
\begin{equation}
\ham(\lambda)=\ham+\lambda\brk*{
\ham^\text{bulk}_\lambda+\ham_{2r,\lambda,\Left}^\bdr+\ham_{2r,\lambda,\Right}^\bdr} +\order{\lambda^2},
\label{eq:FirstorderDeformationH}
\end{equation}
with
\begin{align}
\ham^\text{bulk}_\lambda
&=0,
&
\ham_{2r,\lambda,\Left}^\bdr
&=
-\currentdens_{2r+1}(s_\Left),
&
\ham_{2r,\lambda,\Right}^\bdr
&=
-\currentdens_{2r+1}(s_\Right).
\end{align}
The fact that these deformations merely act on the boundary is in agreement with the observation that they only deform the boundary scattering matrix as shown in the following.
%

%%%%%%%%%%%%%%%%%%%%%%%%%%%%%%%%%%%%%%%%%%%%%%%%%%%%%%%%%%%%%%

%%%%%%%%%%%%%%%%%%%%%%%%%%%%%%%%%%%%%%%%%%%%%%%%%%%
\section{Deformations of Scattering Factors}
\label{sec:OpenChannelScatteringDefs}
%%%%%%%%%%%%%%%%%%%%%%%%%%%%%%%%%%%%%%%%%%%%%%%%%%%
In this section we review the deformations of the scattering factors closely following \cite{Bargheer:2009xy} and \cite{Loebbert:2012yd}. These scattering factors are induced by the above deformations of the conserved charges. We emphasize that essentially the same derivation applies in the context of field theory and lattice models.

%%%%
\subsection{Bulk Scattering Phase}

Consider the two-particle scattering state
\begin{equation}
\ket{u,u'}\simeq a(u,u')\ket{u<u'}+a(u',u)\ket{u'<u},
\label{eq:asymtwopart}
\end{equation}
which is an asymptotic eigenstate of the Hamiltonian:
\begin{equation}
H\ket{u,u'}=
\brk*{h(u)+h(u')}\ket{u,u'}.
\label{eq:bulkeigen}
\end{equation}
Here $\ket{u<u'}$ represents a partial momentum eigenstate (as opposed to the in/out states in \eqref{eqn-inoutstate}) with the particle with rapidity $u(p)$ (or momentum $p$) being on the left of the particle with rapidity $u'(p')$ (or momentum $p'$):
\begin{equation}
\ket{p<p'}=\int_{x\ll x'} e^{\ii px+\ii p'x'} \ket{x,x'},
\end{equation}
The $\simeq$ in \eqref{eq:asymtwopart} signals that we ignore contributions where both particles forming the two-particle state are close to each other. Such contributions will not affect the two-particle scattering factor which is defined as
\begin{equation}
S(u,u')=\frac{a(u',u)}{a(u,u')}.
\end{equation}
We deform the Hamiltonian via the equation
\begin{equation}
\label{eq:defHdef}
\frac{\dd H_\lambda}{\dd\lambda}=\comm{\genX}{H_\lambda},
\end{equation}
where
\begin{equation}
\genX\ket{u,u'}=g(u,u')\ket{u,u'},
\end{equation}
for some eigenvalue function $g$; for the moment we do not specify $g$ or $\genX$, but we will do so in due course.
Differentiating the eigenvalue equation \eqref{eq:bulkeigen} with respect to $\lambda$ and using
\begin{equation}
\frac{\dd}{\dd\lambda}h(u)=0,
\qquad\qquad
\frac{\dd}{\dd\lambda}\ket{u<u'}=0,
\end{equation}
we obtain
\begin{align}
0&=\frac{\dd}{\dd\lambda}\brk[s]*{H_\lambda-h(u)-h(u')}\ket{u,u'}
\nonumber\\
&=
\comm{\genX}{H_\lambda}\ket{u,u'}
+
\brk[s]*{H_\lambda-h(u)-h(u')}\brk*{
\frac{\dd a(u,u')}{\dd\lambda}\ket{u<u'}
+
\frac{\dd a(u',u)}{\dd\lambda}\ket{u'<u}
}
\nonumber\\
&=
\brk[s]*{H_\lambda-h(u)-h(u')}
\brk*{-\genX\ket{u,u'}
+
\frac{\dd a(u,u')}{\dd \lambda}\ket{u<u'}
+
\frac{\dd a(u',u)}{\dd \lambda}\ket{u'<u}
}.
\end{align}
Reading off the coefficients of $\ket{u<u'}$ and $\ket{u'<u}$ yields the equation
\begin{align}
0&=
-g(u,u')a(u,u')+\frac{\dd a(u,u')}{\dd\lambda},
\end{align}
which is solved by
\begin{align}
a(u,u')=e^{\lambda g(u,u')}a_0(u,u').
\end{align}
Hence, the deformed two-particle scattering factor reads
\begin{equation}
S_\lambda(u,u')=e^{\lambda (g(u',u)-g(u,u'))}S(u,u').
\end{equation}
Now we may specify the deformation generator $\genX$ to the bilocal charge operator, such that
\begin{align}
\genX&=\biloc{Q_{r}}{Q_{s}},
&
g(u,u')&=iq_{r}(u)q_{s}(u')+f_{rs}(u)+f_{rs}(u').
\end{align}
Here $f_{rs}$ denotes a local contribution that originates from both constitutent charges of the bilocal operator acting on the same particle.
Hence, we find the following deformation of the bulk scattering matrix:
\begin{equation}
S_\lambda(u,u')=e^{-i\lambda (q_{r}(u)q_{s}(u')-q_{s}(u)q_{r}(u'))}S(u,u').
\label{eq:bilocaldeformationbulkS}
\end{equation}

%%%%
\subsection{Boundary Scattering Phase}

Consider the left boundary scattering state%
\footnote{An analogous investigation applies to the right boundary.}
\begin{equation}
\ket{u}_\Left \simeq a(u)\ket{u}+a(-u)\ket{-u},
\end{equation}
which is an eigenstate of the Hamiltonian:
\begin{equation}
H\ket{u}_\Left=
h(u)\ket{u}_\Left.
\label{eq:leftboundeigen}
\end{equation}
Similarly as above we ignore contributions to the boundary scattering state for which the particle is close to the boundary and which do not affect the scattering factors.
The boundary scattering factor is defined as
\begin{equation}
S_\Left(u)=\frac{a(u)}{a(-u)}.
\end{equation}
We deform the Hamiltonian via the equation
\begin{equation}
\label{eq:defdefX}
\frac{\dd H_\lambda}{\dd\lambda}=\comm{\genX}{H_\lambda},
\end{equation}
where
\begin{equation}
\genX\ket{u}=f(u)\ket{u},
\end{equation}
for some eigenvalue function $f$ and again, for the moment we do not specify $f$ or $\genX$.
Differentiating the eigenvalue equation \eqref{eq:leftboundeigen} with respect to $\lambda$ and using
\begin{equation}
\frac{\dd}{\dd\lambda}h(u)=0,
\qquad\qquad
\frac{\dd}{\dd\lambda}\ket{u}=\frac{\dd}{\dd\lambda}\ket{-u}=0,
\end{equation}
we obtain
\begin{align}
0&=\frac{\dd}{\dd\lambda}\brk[s]*{H_\lambda-h(u)}\ket{u}_\Left
\nonumber\\
&=
\comm{\genX}{H_\lambda}\ket{u}_\Left
+
\brk[s]*{H_\lambda-h(u)}\brk*{
\frac{\dd a(u)}{d\lambda}\ket{u}
+
\frac{\dd a(-u)}{d\lambda}\ket{-u}
}
\nonumber\\
&=
\brk[s]*{H_\lambda-h(u)}
\brk*{-\genX\ket{u}_\Left
+
\frac{\dd a(u)}{\dd\lambda}\ket{u}
+
\frac{\dd a(-u)}{\dd\lambda}\ket{-u}
}.
\end{align}
Reading off the coefficients of $\ket{u}$ and $\ket{-u}$ yields the equations
\begin{align}
0&=
-f(u)a(u)+\frac{\dd a(u)}{\dd\lambda},
\nonumber\\
0&=
-f(-u)a(-u)+\frac{\dd a(-u)}{\dd\lambda},
\end{align}
which are solved by
\begin{align}
a(u)=e^{\lambda f(u)}a_0(u).
\end{align}
Hence, the deformed boundary scattering factor reads
\begin{equation}
S_{\Left,\lambda}(u)=e^{\lambda (f(u)-f(-u))}S_\Left(u).
\label{eq:DeformedSOddCharges}
\end{equation}
Now we may specify the deformation generator $\genX$ to one of the two cases which induce deformations of the left boundary scattering matrix:
\begin{align}
&1):\quad \genX=\biloc{Q_{2r}}{Q_{2s+1}},
&
f(u)&=\ihalf q_{2r}(u)q_{2s+1}(u),
\label{eq:boundSdefbiloc}
\\
&2):\quad \genX=Q_{2r+1},
&
f(u)&=iq_{2r+1}(u).
\label{eq:boundSdefloc}
\end{align}
Note that the factor $1/2$ in \eqref{eq:boundSdefbiloc} originates from the $1/2$ in front of the local contribution $q_{2r}(x)q_{2s+1}(x)$ to the bilocal operator as prescribed by the definition \eqref{eq:bilocfield}.
Analogously we can proceed for the right boundary where we use $\genX=\biloc{Q_{2s+1}}{Q_{2r}}$.

For deformations generated by $ \genX=\biloc{Q_{2r}}{Q_{2s+1}}$ the deformed bulk S-matrix takes the form (cf.\ \eqref{eq:bilocaldeformationbulkS})
\begin{equation}
S_\lambda(u,u')=e^{-i\lambda (q_{2r}(u)q_{2s+1}(u')-q_{2s+1}(u)q_{2r}(u'))}S(u,u'),
\end{equation}
and the boundary scattering factor deformed by \eqref{eq:boundSdefbiloc} we have
\begin{align}
S_{\Left,\lambda}(u)=S_\lambda(u,-u)S_{\Left,\lambda}(-u),
\end{align}
which is the boundary cross-unitarity condition of  \cite{Ghoshal:1993tm}.

%%%%%%%

\subsection{Defect Scattering Phase}
The derivation of the deformed scattering phase parallels the boundary case. We first consider the
topological case and then the non-topological one.

\paragraph{Topological Defects.}
The topological defect is purely transmissive. To determine the deformed transmissive amplitude, we consider the following one-particle state
\begin{align}
|u\rangle_{\rD}=a(u)|u;\varnothing\rangle+b(u)|\varnothing;u\rangle.
\end{align}
The transmissive amplitude is given by
\begin{align}
T_-(u)=T_+(u)=T(u)=\frac{b(u)}{a(u)}.
\end{align}
Consider the bilocal deformation (\ref{eq:defdefX}) of the Hamiltonian. We denote the deformed Hamiltonian by $H_{\lambda}$. In the infinite volume limit, we have
\begin{align}
\label{eq:eigendefect}
H_{\lambda}|u\rangle_{\rD}=h(u)|u\rangle_{\rD}.
\end{align}
As before, the asymptotic states diagonalize the operator $X$:
\begin{align}
X|u;\varnothing\rangle=f(u)|u;\varnothing\rangle,\qquad X|\varnothing;u\rangle=f(u)|\varnothing;u\rangle.
\end{align}
Taking the derivative of (\ref{eq:eigendefect}) with respect to $\lambda$, we obtain
\begin{align}
\label{eq:XH}
[X,H_{\lambda}]|u\rangle_{\rD}+H_{\lambda}\frac{\dd}{\dd\lambda}|u\rangle_{\rD}=h(u)\frac{\dd}{\dd\lambda}|u\rangle_{\rD}
\end{align}
where we have used the fact that
\begin{align}
\frac{\dd}{\dd\lambda}|u;\varnothing\rangle=\frac{\dd}{\dd\lambda}|\varnothing;u\rangle=0,\qquad \frac{\dd}{\dd\lambda}h(u)=0.
\end{align}
Equation (\ref{eq:XH}) can be brought to the form
\begin{align}
[H_{\lambda}-h(u)]\left(-X|u\rangle_{\rD}+\frac{\dd a(u)}{\dd\lambda}|u;\varnothing\rangle+\frac{\dd b(u)}{\dd\lambda}|\varnothing;u\rangle\right)=0,
\end{align}
which implies
\begin{align}
-f(u)+\frac{\dd a(u)}{\dd\lambda}=0,\qquad -f(u)+\frac{\dd b(u)}{\dd\lambda}=0.
\end{align}
These equations can be solved by
\begin{align}
a_{\lambda}(u)=e^{\lambda f(u)}a(u),\qquad b_{\lambda}(u)=e^{\lambda f(u)}b(u).
\end{align}
This leads to the conclusion that the transmission amplitude is not affected:
\begin{align}
T_{\lambda}(u)=\frac{a_{\lambda}(u)}{b_{\lambda}(u)}=\frac{a(u)}{b(u)}=T(u).
\end{align}
Physically, this is expected for the $T\bar{T}$ deformation. Topological defects, by definition, are invariant under variations of the metric. As a result, they are not sensitive to the stress energy tensor. Since the $T\bar{T}$ operator is constructed from the stress energy tensor, it is natural that topological defects are not affected by the $T\bar{T}$ deformation. From our derivation, we see that the topological defect is not affected by these deformations.

\paragraph{Non-Topological Defects.} Integrable defects that are non-topological are only allowed in free theories with the bulk S-matrix being $S=\pm 1$. In this case, the one-particle state is given by
\begin{align}
|u\rangle_{\rD}=a(u)|u;\varnothing\rangle+b(u)|\varnothing;u\rangle+c(u)|-u;\varnothing\rangle.
\end{align}
The transmission and reflection amplitudes are given by
\begin{align}
T(u)=\frac{b(u)}{a(u)},\qquad R(u)=\frac{c(u)}{a(u)}.
\end{align}
Going through the same steps, we find that the corresponding deformed quantities are
\begin{align}
R_{\lambda}(u)=e^{\lambda f(u)-\lambda f(-u)}R(u),\qquad T_{\lambda}(u)=T(u).
\end{align}
We see that the reflection amplitude is deformed in the same way as in the boundary case while the transmission amplitude is undeformed like the topological defect. Here we note similarities to the deformation factor obtained in \cite{Serban:2013jua}.

%%%%%%%%%%%%%%%%%%%%%%%%%%%%%%%%%%%%%%%%%%%%%%%%%%%%%%%%%%%%%%
\section{Finite Volume Spectrum}
\label{sec:spectrum}
\begin{figure}[t]
\centering
\includegraphics[scale=0.5]{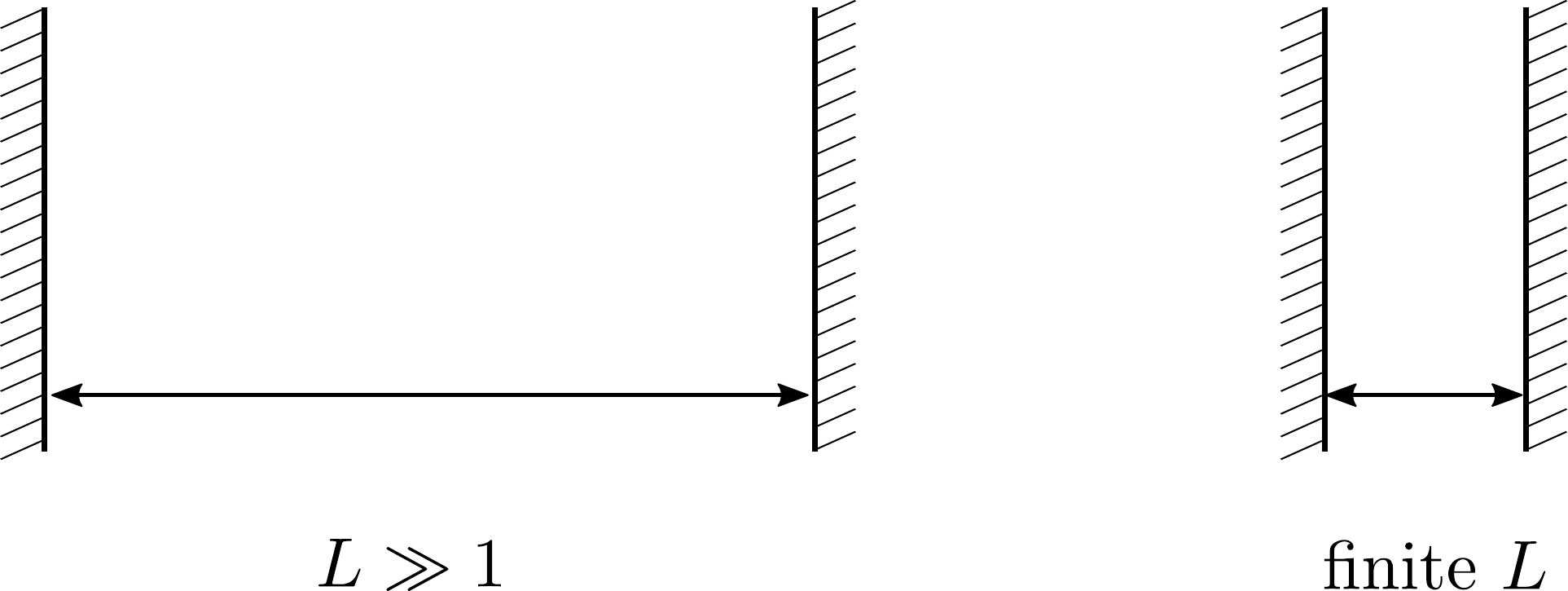}
\caption{Theories defined on a strip.}
\label{fig:boundary}
\end{figure}

%%%%%%%%%%%%%%%%%%%%%%%%%%%%%%%%%%%%%%%%%%%%%%%%%%%%%%%%%%%%%%
In the previous sections, we have derived the deformed bulk and boundary S-matrices. This data allows us to compute important physical quantities. In this section, we will focus on the finite volume spectrum for theories defined on a strip,
 \emph{i.e.}\ with two integrable boundaries in the spatial direction, as is shown in \figref{fig:boundary}.
In the left panel of \figref{fig:boundary}, the distance between the boundaries $L$ is large such that the spectrum can be well approximated by the asymptotic Bethe ansatz; in the right panel, the distance $L$ is finite and the spectrum needs to be determined by the boundary thermodynamic Bethe ansatz (BTBA).\par

We will consider the deformed spectrum for two types of deformations. The first one is the bilinear deformation triggered by the operators
\begin{align}
\mathcal{O}_{rs}=\varepsilon_{\mu\nu}J_{P_r}^{\mu}J_{H_s}^{\nu},
\end{align}
or alternatively by the bilocal operators
\begin{equation}
\biloc{\mom_r}{\ham_s} \quad \text{or } \quad \biloc{\ham_s}{\mom_r},
\end{equation}
respectively (see \secref{sec:DefsClosedBounds} for relation between the two alternative pictures).
Among the general bilinear deformations, two families are of special interest. The case $r=s$ denotes the Castillejo-Dalitz-Dyson (CDD) deformations. These are the deformations that \emph{preserve Lorentz invariance}. The additional phase factor inherited by the deformed S-matrix is the famous CDD factor. The case $r=1$ is also special and shall be called the dynamical hard rod deformation. This family is interesting because it has the nice physical interpretation to correspond to deforming point particles into finite size hard rods. The two families of deformations coincide for $s=1$, which is the $T\bar{T}$ deformation. Let us comment on the possible values of the index $s$ of higher charges $P_s,H_s$. A priori, $s$ runs over the odd integers. However, for a given theory, in general $s$ does not cover all odd values. The possible set of values of $s$ is an important characteristic of the model, see \cite{Zamolodchikov:1989hfa} for more details. Since we aim at developing the general framework and do not specify our considerations to a certain theory, we will simply take $s$ to be an odd number.\par

The other type of deformation, which is specific to the boundary case, is triggered by an odd charge and defined by
\begin{align}
\frac{\rd H_{\lambda}}{\rd\lambda}=[P_r,H_{\lambda}].
\end{align}
This type of deformations does not change the bulk S-matrix (\ref{eq:boundSdefloc}), but has non-trivial effects on the boundary S-matrices, which deform the spectrum in an interesting way.

%%%%%%%%%%%%%%

\subsection{Large Volume Limit}
We first consider the limit $L\gg 1$. In this limit, the spectrum is determined by the asymptotic Bethe equations
\begin{align}
\label{eq:BAE}
e^{2i p(u_j)L}S_{\rL}(u_j)S_{\rR}(-u_j)\prod_{k\ne j}^NS(u_j,u_k)S(u_j,-u_k)=1,\qquad j=1,\cdots,N.
\end{align}
where $S(u,v)$ is the bulk S-matrix and $S_{\rL,\rR}(u)$ denotes the left and right boundary S-matrix, respectively.

\begin{figure}[t]
\centering
\includegraphics[scale=0.18]{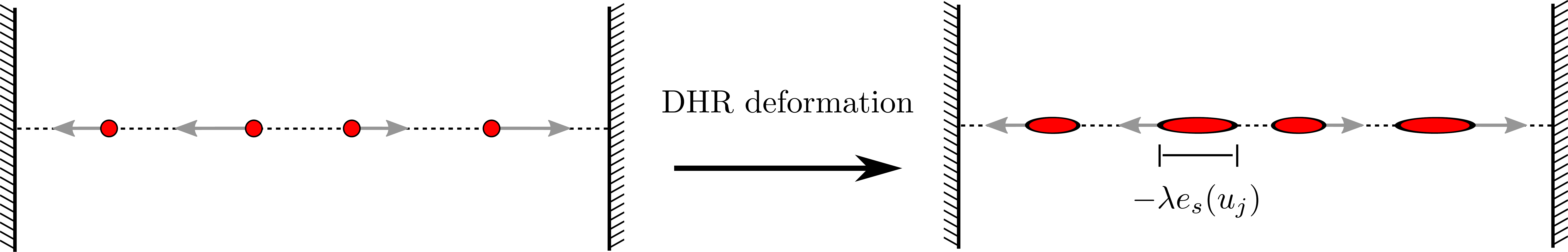}
\caption{The dynamical hard rod interpretation for the $\mathcal{O}_{1s}$ deformation.}
\label{fig:DHR}
\end{figure}

\paragraph{Bilinear Deformations.} For the bilinear deformations, the deformed bulk and boundary matrices are given by\footnote{See \cite{Beisert:2008cf,Loebbert:2012yd} for similar results in the spin chain case.} \eqref{eq:bilocaldeformationbulkS} and \eqref{eq:DeformedSOddCharges}, which we quote here
\begin{align}
\label{eqn-DeformedS}
S_{\lambda}(u,v)=&\,S(u,v)e^{-i\lambda(p_r(u)e_s(v)-e_s(u)p_r(v))},\\\nonumber
S_{\rL,\lambda}(u)=&\,S_{\rL}(u)e^{i\lambda p_r(u)e_s(u)},\\\nonumber
S_{\rR,\lambda}(u)=&\,S_{\rR}(u)e^{-i\lambda p_r(u)e_s(u)},
\end{align}
where
\begin{align}
p_r(u)=\gamma_r\sinh(r u),\qquad e_s(u)=\gamma_s \cosh(s u),
\end{align}
and both $r,s$ are odd integers. We have
\begin{align}
p_r(-u)=-p_r(u),\qquad e_s(-u)=e_s(u).
\end{align}
For deformations involving higher charges, it is more convenient to consider the twisted Bethe equations
\begin{align}
e^{2iL(p(u_j)+\nu_r p_r(u_j))}S_{\rL}(u_j)S_{\rR}(-u_j)\prod_{k\ne j}^N S(u_j,u_k)S(u_j,-u_k)=1,
\end{align}
where $\nu_r$ is the twist that couples to the
odd ($\mom$-type) charges. The deformed Bethe equations read
\begin{align}
e^{2iL(p(u_j)+\nu_r p_r(u_j))}\prod_{k=1}^N e^{2i\lambda p_r(u_j)e_s(u_k)}\,S_{\rL}(u_j)S_{\rR}(-u_j)\prod_{k\ne j}^N S(u_j,u_k)S(u_j,-u_k)=1.
\end{align}
They can be recast as
\begin{align}
e^{2iL(p(u_j)+\nu_r p_r(u_j))}\,S_{\rL}(u_j)S_{\rR}(-u_j)\prod_{k\ne j}^N S(u_j,u_k)S(u_j,-u_k)=e^{-2i\lambda Q_N^{(s)}\,p_r(u_j)},
\end{align}
where
\begin{align}
Q_N^{(s)}=\sum_{k=1}^N e_s(u_k),
\qquad
E_N=Q_N^{(1)}.
\end{align}
We see that the deformed Bethe equations take the same form as the undeformed ones, except that the twist $\nu_r$ is shifted according to
\begin{align}
\nu_r\to \nu_r+\frac{\lambda Q_N^{(s)}}{L}.
\end{align}
This implies the following flow equation for the energy and total even charge:
\begin{align}
\label{eq:flowlargeR0}
\partial_{\lambda}E_N=\frac{1}{L}Q_N^{(s)}\partial_{\nu_r}E_N,\qquad \partial_{\lambda}Q_N^{(s)}=\frac{1}{L}Q_N^{(s)}\partial_{\nu_r}Q_N^{(s)}.
\end{align}
The case $r=1$ is special. In this case $p_r(u_j)$ coincides with the momentum and there is no need to introduce additional twist. The deformation has the effect of changing the length $L$ as
\begin{align}
L\to L+\lambda Q_N^{(s)}.
\end{align}
This leads to the following flow equations
\begin{align}
\label{eq:flowlargeR}
\partial_{\lambda}E_N=Q_N^{(s)}\partial_L E_N,\qquad Q_N^{(s)}=Q_N^{(s)}\partial_L Q_N^{(s)}.
\end{align}
The deformations triggered by $\mathcal{O}_{1s}$ have an interesting intuitive interpretation as is shown in \figref{fig:DHR}.
For $\lambda<0$, the deformation turns point particles into finite size hard rods with length $-\lambda e_s(u_j)$. Therefore, the effective length, which describes the `free space' between the rods, is reduced and becomes $L+\lambda Q_N^{(s)}$. For $\lambda>0$, the distances between the particles are increased and the effective length become larger. This interpretation was first proposed in non-relativistic models \cite{Cardy:2020olv,Jiang:2020nnb,Hansen:2020hrs}. Here we see a natural generalization to the relativistic case.

\begin{figure}[t]
\centering
\includegraphics[scale=0.18]{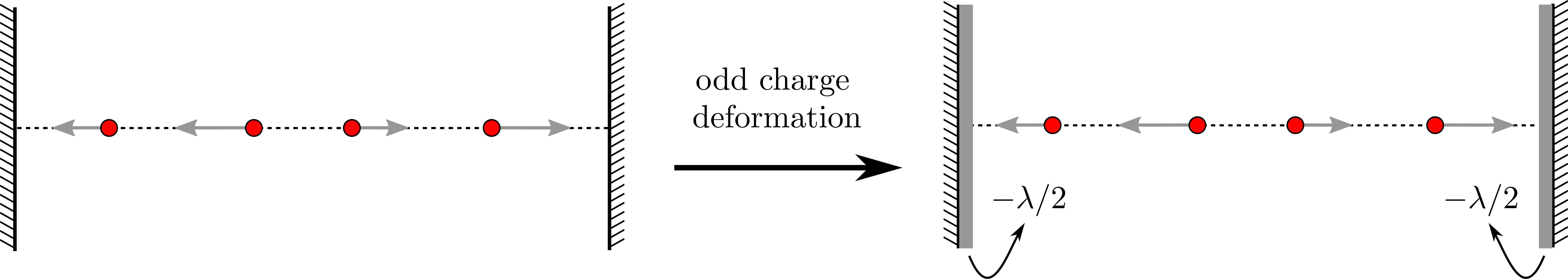}
\caption{The `thick wall' interpretation of the odd charge deformation.}
\label{fig:odd}
\end{figure}

\paragraph{The Odd Charge Deformations.} For the odd charge deformations triggered by $P_r$, the deformed bulk and boundary S-matrices read
\begin{align}
\label{eqn-DeformedSOdd}
S_{\lambda}(u,v)=&\,S(u,v),\\\nonumber
S_{\rL,\lambda}(u)=&\,S_{\rL}(u)e^{i\lambda p_r(u)},\\\nonumber
S_{\rR,\lambda}(u)=&\,S_{\rR}(u)e^{-i\lambda p_r(u)}.
\end{align}
Notice that the bulk S-matrix is undeformed. The deformed asymptotic Bethe equation takes the following form
\begin{align}
e^{2iL(p(u_j)+\nu_r p_r(u_j))}\,S_{\rL}(u_j)S_{\rR}(-u_j)\prod_{k\ne j}^N S(u_j,u_k)S(u_j,-u_k)=e^{-2i\lambda p_r(u_j)}.
\end{align}
We see that it can be brought to the original form by shifting the chemical potential
\begin{align}
\nu_r\to \nu_r+\frac{\lambda}{L}.
\end{align}
This implies the following flow equation for the energy and total even charge:
\begin{align}
\label{eq:oddflow0}
\partial_{\lambda}E_N=\frac{1}{L}\partial_{\nu_r}E_N.
\end{align}
The case $r=1$ is again special. The deformed Bethe equation is simply obtained from the undeformed one by setting
\begin{align}
L\to L+\lambda,
\end{align}
which implies a linear flow equation for the spectrum:
\begin{align}
\label{eq:oddflow1}
\partial_{\lambda}E_N=\partial_LE_N.
\end{align}
We also have an intuitive interpretation for this result, as is shown in \figref{fig:odd}. For $\lambda<0$, the deformation make the boundary `thicker', which reduces the distances between the two boundaries by $|\lambda|$.
For $\lambda>0$, the distance between the boundaries is increased by $|\lambda|$.\par

We make one comment before ending this subsection. Although the flow equations we have obtained so far are in the large volume limit, we expect that they should hold also in the finite volume, based on the experience on previous results in the bulk case. As we shall see, this will be confirmed by the boundary TBA computation below.

\subsection{Finite Volume}
Now we consider the situation where $L$ is finite. Due to finite size corrections, the asymptotic Bethe ansatz is no longer sufficient. To obtain the spectrum, we exploit the boundary TBA approach. The idea of this approach is to translate the calculation of the finite size spectrum to the calculation of the thermal free energy of the \emph{mirror theory}. The mirror theory is defined in the infinite volume limit and the asymptotic description is valid. Because our deformations involve higher conserved charges, we need to consider a generalized partition function that contains additional chemical potentials and charges. In the mirror theory, these chemical potentials correspond to \emph{twists}, which enter the quantization conditions of the mirror rapidities \cite{Hernandez-Chifflet:2019sua}.

%%%%%%%%

\paragraph{Double Wick Rotation.} The mirror theory is obtained from the physical theory by performing a double Wick rotation which swaps the role of space and time:
\begin{align}
H\mapsto i\widetilde{P},\qquad P\mapsto i\widetilde{H}.
\end{align}
Here we use a tilde to denote quantities in the mirror theory. Under the double Wick rotation, the rapidity is transformed as $u\mapsto u+\tfrac{i\pi}{2}$. For higher charges with $s=2r-1$, we have similarly
\begin{align}
\label{eq:mirrorhighH}
H_{2r-1}\mapsto i(-1)^{r-1}\widetilde{P}_{2r-1},\qquad P_{2r-1}\mapsto i(-1)^{r-1}\widetilde{H}_{2r-1},\qquad r=1,2,\dots.
\end{align}
The single-particle eigenvalues of the higher mirror charges $\widetilde{P}_{2r-1}$, $\widetilde{H}_{2r-1}$ take the same form as in the original theory
\begin{align}
\tilde{e}_{2r-1}(u)=\gamma_{2r-1}\cos(2r-1)u,\qquad \tilde{p}_{2r-1}=\gamma_{2r-1}\sin(2r-1)u.
\end{align}
The mirror bulk S-matrix is given by the mirror transformation of the physical S-matrix. The boundary S-matrix is related to the two-particle form factor $K(u)$ in the mirror channel. More explicitly, we have
\begin{align}
\widetilde{S}(u,v)=S\left(u+\tfrac{i\pi}{2},v+\tfrac{i\pi}{2}\right),\qquad K(u)=S_{\rR}\left(u-\tfrac{i\pi}{2}\right).
\end{align}

\begin{figure}[t]
\centering
\includegraphics[scale=0.45]{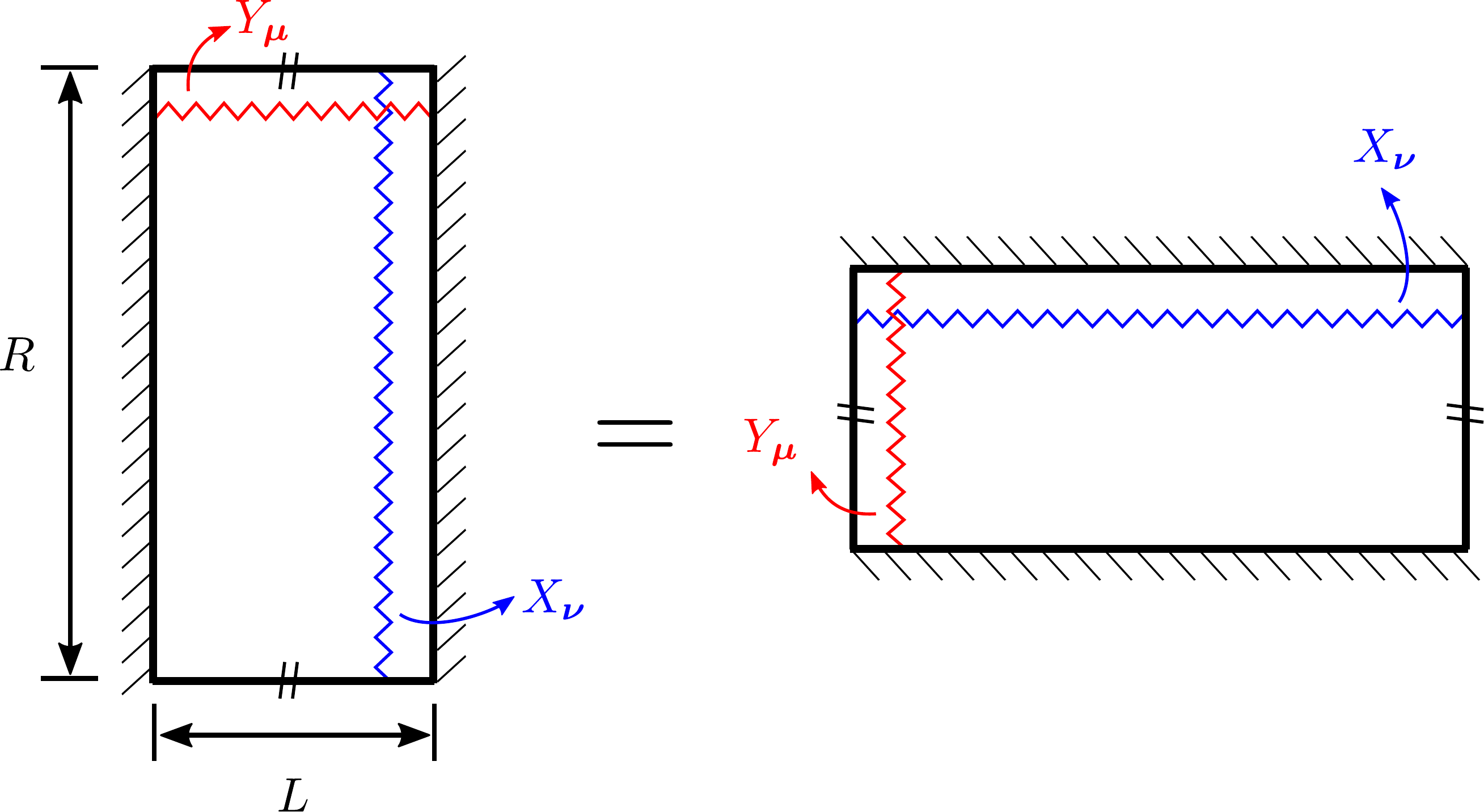}
\caption{Computing the ground state vacuum using the mirror TBA. We introduce twists and chemical potentials, which get mapped into each other under the mirror transformation.}
\label{fig:twists}
\end{figure}

\paragraph{Generalized Mirror BTBA.} Now we consider the generalized mirror boundary TBA. In the physical channel, the length of the space is $L$. We take the length of the periodic direction to be $R$. The generalized partition function takes the following form:
\begin{align}
\label{eq:Zphysical}
Z_{ab}=\tr\left[e^{-R(H+\mu_s H_{s})} \right].
\end{align}
Here we have introduced the additional chemical potentials $\mu_s$ for the even charges.%
\footnote{In principle we could also introduce chemical potentials for the odd charges $P$ and $P_r$. However, for a theory with integrable boundaries the odd charges are zero and we can set the chemical potential to zero from the beginning.}
In the limit $R\gg 1$, the partition function is dominated by the ground state charges:
\begin{align}
\label{eq:largeR1}
Z_{ab}\sim e^{-R\left[E^{(0)}(L)+\mu_s E_{s}^{(0)}(L)\right]}.
\end{align}
To compute the energy and the higher charge of the ground state, we go to the mirror kinematics, as is shown in \figref{fig:twists}.
The partition function in the mirror kinematics takes the form
\begin{align}
\label{eq:Zmirror}
Z_{ab}=\langle B_a|e^{-L(\widetilde{H}+(-1)^{\frac{r+1}{2}}\nu_r\widetilde{H}_r)}|B_b\rangle,
\end{align}
where $|B_{j}\rangle$ ($j=a,b$) are the boundary states which correspond to the two boundaries in the open channel. We introduced an additional $(-1)^{(r+1)/2}$ in the definition of the chemical potential for later convenience.
The chemical potential in (\ref{eq:Zphysical}) is implicitly contained in the mirror partition function. It becomes the \emph{twist} in the mirror channel, which enters the quantization condition for the mirror rapidities. The mirror partition function (\ref{eq:Zmirror}) can be written as
\begin{align}
Z_{ab}=\sum_{n}\frac{\langle B_a|n\rangle\langle n|B_b\rangle}{\langle n|n\rangle}e^{-L\mathcal{X}_n(R)},
\end{align}
where the sum is over all eigenstates of the Hamiltonian and $\mathcal{X}_n(R)$ is defined as
\begin{align}
\left(\widetilde{H}+(-1)^{\frac{r+1}{2}}\nu_r\widetilde{H}_r\right)|n\rangle=\mathcal{X}_n(R)|n\rangle.
\end{align}
For integrable boundaries, the overlap $\langle B_a|n\rangle$ is only non-zero for the states with \emph{paired rapidities}
\begin{align}
|n\rangle=|\alpha_{2N}\rangle=|u_N,-u_N,\cdots,u_1,-u_1\rangle,
\end{align}
where $u_N>u_{N-1}>\cdots>u_1$. For a state with paired rapidities $\{u_1,-u_1,\cdots,u_N,-u_N\}$, the eigenvalue is given by
\begin{align}
\mathcal{X}_n(R)=2\sum_{j=1}^N X_{{\nu}}(u_j),
\end{align}
where
\begin{align}
X_{{\nu}}(u)=\tilde{e}(u)+(-1)^{\frac{r+1}{2}}\nu_r\tilde{e}_r(u).
\end{align}
In the mirror channel for $R\gg 1$, the spectrum can be described by the asymptotic Bethe ansatz. The rapidities satisfy the mirror Bethe ansatz equations. In the usual case, the mirror Bethe ansatz equations take the following form
\begin{align}
\label{eq:mBAEtwist}
e^{i m R\sinh(u_i)}\widetilde{S}(u_i,-u_i)\prod_{j\ne i}^N\widetilde{S}(u_i,u_j)\widetilde{S}(u_i,-u_j)=1.
\end{align}
As mentioned before, we have additional twists which come from the chemical potential in the physical channel. In this case, we need to modify the Bethe equations by the replacement
\begin{align}
e^{imR\sinh(u)}\mapsto e^{R Y_{\bm{\mu}}\left(u+\frac{i\pi}{2}\right)},
\end{align}
where
\begin{align}
Y_{\bm{\mu}}(u)=e(u)+\mu_se_s(u).
\end{align}
Following the standard procedure, we introduce the density of pair rapdities and holes $\tilde{\rho}(u)$, $\tilde{\rho}_h(u)$. From (\ref{eq:mBAEtwist}), we have
\begin{align}
\tilde{\rho}(u)+\tilde{\rho}_h(u)=\frac{\partial_uY_{\bm{\mu}}\left(u+\frac{i\pi}{2}\right)}{2\pi i}+(\widetilde{\varphi}_+*\tilde{\rho})(u),
\end{align}
where
\begin{align}
\widetilde{\varphi}(u,v)=\frac{1}{2\pi i}\frac{\partial}{\partial u}\log \widetilde{S}(u,v),
\end{align}
and
\begin{align}
\widetilde{\varphi}_+(u,v)=\widetilde{\varphi}(u,v)+\widetilde{\varphi}(u,-v),\qquad
(\widetilde{\varphi}_{+}*\tilde{\rho})(u)=\int_0^{\infty}\widetilde{\varphi}(u,v)\tilde{\rho}(v)\rd v.
\end{align}
Therefore the partition function can be written as
\begin{align}
\label{eq:funcZab}
Z_{ab}=\int\mathcal{D}\tilde{\rho}\,\exp\left[R\int_0^{\infty}\left(\log[\chi_{ab}(u)]-2LX_{{\nu}}(u) \right)\tilde{\rho}(u)\rd u+S[\tilde{\rho},\tilde{\rho}_h] \right],
\end{align}
where
\begin{align}
\chi_{ab}(u)=\overline{K}_a(u)K_b(u),
\end{align}
and $S[\tilde{\rho},\tilde{\rho}_h]$ is the Yang-Yang entropy
\begin{align}
S[\tilde{\rho},\tilde{\rho}_h]=R\int_0^{\infty}\left[ (\tilde{\rho}+\tilde{\rho}_h)\log(\tilde{\rho}+\tilde{\rho}_h)-\tilde{\rho}\log\tilde{\rho}-\tilde{\rho}_h\log\tilde{\rho}_h\right]\rd u.
\end{align}
In the limit $R\to\infty$, the partition function $Z_{ab}$ (\ref{eq:funcZab}) is dominated by the saddle point. The saddle-point equation is the boundary TBA equation
\begin{align}
\label{eq:BTAB}
\epsilon(u)=2LX_{{\nu}}(u)-\log[\chi_{ab}(u)]-\log(1+e^{-\epsilon})*\varphi_+,
\end{align}
where $\epsilon(u)$ is the pseudo-energy given by $e^{\epsilon(u)}=\tilde{\rho}_h(u)/\tilde{\rho}(u)$.
The free energy reads
\begin{align}
F=\frac{R}{2\pi i}\int_0^{\infty}\partial_uY_{\bm{\mu}}\left(u+\tfrac{i\pi}{2}\right)\,\log\left(1+e^{-\epsilon(u)}\right)\rd u.
\end{align}
Comparing with (\ref{eq:largeR1}), we obtain the expressions for the finite volume charges in the ground state:
\begin{align}
E^{(0)}(L,\nu_r)=&\,-\frac{1}{2\pi i}\int_0^{\infty}\partial_u e(u+i\pi/2)\,\log\left(1+e^{-\epsilon(u)}\right)\rd u,\\\nonumber
Q_{s}^{(0)}(L,\nu_r)=&\,-\frac{1}{2\pi i}\int_0^{\infty}\partial_u e_s(u+i\pi/2)\,\log\left(1+e^{-\epsilon(u)}\right)\rd u.\\\nonumber
\end{align}
The deformed TBA kernel is given by
\begin{align}
\tilde{\varphi}_{+,\lambda}(u,v)=&\,\tilde{\varphi}_+(u,v)+\frac{\lambda}{\pi}\partial_u e_s(u+\tfrac{i\pi}{2})p_r(v+\tfrac{i\pi}{2})\\\nonumber
=&\,\tilde{\varphi}_+(u,v)-(-1)^{\frac{r-1}{2}}\frac{\lambda}{\pi i}\partial_u e_s(u+\tfrac{i\pi}{2})\tilde{e}_r(v).
\end{align}
Plugging this into the TBA equation (\ref{eq:BTAB}), we find that it takes the same form as the undeformed one, except for the shift of $\nu_r$ according to
\begin{align}
\nu_r\to \nu_r+\frac{\lambda}{L}Q_s^{(0)}.
\end{align}
This implies the following flow equation for the ground state energy and charge:
\begin{align}
\label{eq:flowF0}
\partial_{\lambda} E_{\lambda}^{(0)}=\frac{1}{L}Q_{s,\lambda}^{(0)}\partial_{\nu_r}E_{\lambda}^{(0)},\qquad
\partial_{\lambda}Q_{s,\lambda}^{(0)}=\frac{1}{L}Q_{s,\lambda}^{(0)}\partial_{\nu_r}Q_{s,\lambda}^{(0)}.
\end{align}
This is precisely the same flow equation as we obtained in the large volume limit (\ref{eq:flowlargeR}). Again the case $r=1$ is special. In this case, we simply have
\begin{align}
L\to L+\lambda Q_s^{(0)}
\end{align}
and the flow equation becomes
\begin{align}
\label{eq:flowF1s}
\partial E_{\lambda}^{(0)}=Q_{s,\lambda}^{(0)}\partial_{L}E_{\lambda}^{(0)},\qquad
\partial_{\lambda}Q_{s,\lambda}^{(0)}=Q_{s,\lambda}^{(0)}\partial_{L}Q_{s,\lambda}^{(0)}.
\end{align}
The flow equations \eqref{eq:flowF0} and \eqref{eq:flowF1s} are the same as in the large volume case \eqref{eq:flowlargeR0} and \eqref{eq:flowlargeR}, as expected.\par

We can perform the same analysis for the odd charge deformations in the finite volume. The resulting flow equations of the spectrum again take the same form as the ones in the large volume limit \eqref{eq:oddflow0} and \eqref{eq:oddflow1}.

%%%%%%%%%%%%%%%%%%%%%%%%%%%%%%%%%%%%%%%%%%%%%%%%%%%%%%%%%%%%%%
\section{Deformed Partition Function and Exact $g$-Function}
\label{sec:TorusA}
%%%%%%%%%%%%%%%%%%%%%%%%%%%%%%%%%%%%%%%%%%%%%%%%%%%%%%%%%%%%%%
In this section, we discuss the flow equation for the partition function and the $g$-function, under the $T\bar{T}$ deformation. We shall restrict to the case where the partition function takes the form of the ordinary thermo partition function. For integrable systems, we could have generalized partition functions that also depend on higher charges, see the discussion in \appref{apd-HigherDeform}.

\subsection{Asymptotic Behavior of the Partition Function}

For generic QFTs at thermal equilibrium, the quantum states with energy $E$ are fully characterized by the Boltzmann weight $\exp{(-\beta E)}$. The same also holds for IQFTs with vanishing higher charges, because those charges will stay zero after the deformation.

\paragraph{Asymptotic Limit.}

Before discussing the deformed partition function and the $g$-function, let us first study the general behavior of the partition function in the large volume limit, where the circumference $R$ and the height $L$ of the cylinder are both large. For massive IQFT, the precise meaning of large is $R,L \gg 1/m$, where $m$ is the mass gap of the theory. We remind that  the open and closed channel description was briefly introduced in \secref{sec:openclosed}.

\paragraph{Open Channel.}
In the open channel, the partition function is given by
\begin{align}
\label{eq:openZ}
Z_{ab}(R,L)=\text{Tr}\,e^{-H_{ab}R}=\sum_{\psi}e^{-E_{ab}^{\psi}(L)R},
\end{align}
where $H_{ab}$ is the open channel Hamiltonian and the sum is over all the eigenstates of $H_{ab}$, denoted here by $\psi$.
In the limit $mR\gg 1$, the partition function is dominated by the ground state energy $E_{ab}^{(0)}(L)$. In the large $L$ limit, the ground state energy has the following universal behavior
\begin{align}
\label{eq:E0largeL}
E_{ab}^{(0)}(L)=\epsilon_0\,L+f_a+f_b+\mathcal{O}(e^{-mL}),
\end{align}
where $\epsilon_0$ is the bulk energy density and $f_{a,b}$ are the non-extensive contributions from the boundary. To sum up, in the limit $mR,mL\gg 1$, we find
\begin{align}
\label{eq:largeZopen}
Z_{ab}(R,L)=\,e^{-R(\epsilon_0L+f_a+f_b)}+\cdots =\,e^{-RL\epsilon_0-R f_a-R f_b}+\cdots.
\end{align}

\paragraph{Closed Channel.}
In the closed channel, the partition function is given by
\begin{align}
\label{eq:closedZ}
Z_{ab}(R,L)=\langle B_a|e^{-H(R)L}|B_b\rangle=\sum_{\phi}(G_a^{\phi}(R))^*G_b^{\phi}(R)\,e^{-E_{\phi}(R)L},
\end{align}
where the amplitudes $G_j^{\phi}(R)$ are defined as the normalized overlaps
\begin{align}
G_j^{\phi}(R)=\frac{\langle\phi|B_j\rangle}{\sqrt{\langle\phi|\phi\rangle}},\qquad j=a,b.
\end{align}
In the limit $mL\gg 1$, the partition function is dominated by the ground state
\begin{align}
Z_{ab}(R,L)\sim [G_a^{(0)}(R)]^* G_b^{(0)}(R) e^{-E_0(R)L}.
\end{align}
Taking the $mR\gg 1$ limit of this expression, we find
\begin{align}
E_0(R)\approx\epsilon_0 R+\mathcal{O}(e^{-mR}),
\end{align}
where $\epsilon_0$ is the same quantity as the one which appears in (\ref{eq:E0largeL}).  As opposed to \eqref{eq:E0largeL}, $E_0(R)$ does not have $\mathcal{O}(1)$ contributions from the boundaries, since the system is closed. Therefore, in the $mR, mL \gg 1$ limit, we have
\begin{align}
Z_{ab}(R,L)\sim [G_a^{(0)}(R)]^* G_b^{(0)}(R)e^{-RL\epsilon_0}+\cdots.
\end{align}
%%%%%%%%%%%%%%%%
\paragraph{Definition of the $g$-Function.}

Comparing the asymptotics in the closed channel with the open channel, (\ref{eq:largeZopen}), we must have
\begin{align}
G_j^{(0)}(R)\sim e^{-f_j R}(1+\cdots)
\end{align}
Namely, the overlap should be $e^{-f_jR}$ multiplied by some order 1 quantity. We will define this quantity as the $g$-function:
\begin{align}
\log g_a(R)=\log G_a^{(0)}(R)+f_a R.
\label{eq:gandG}
\end{align}

To summarize, the exact $g$-function, or the boundary entropy, is defined as the overlap between the finite volume vacuum state and the boundary state. To extract important information about the boundary, we have subtracted a universal constant part $f_a R$ from the naive overlap $G_j^{0}$, in the definition of $g$. The exact $g$-function is an important quantity of interest in QFTs with boundaries. It describes the boundary degrees of freedom \cite{Affleck:1991tk} and has interesting properties along RG flow \cite{Friedan:2003yc}.

%%%%%%%%%%%%%%%%%%%%%%%%%%%%%%%%%%%%%%%%%%%%%%%%%%%%%%%

\subsection{Flow Equation}

In this section, we present derivations of the flow equation of the $T\bar{T}$ deformed partition function and the $g$-function.
Our strategy is as follows. The flow equation for the spectrum in the open channel is known from the exact deformed S-matrices.
Since the spectrum are deformed universally, that is, all the spectrum obey the same flow equation, we can write down the equation for the partition function in the open channel. We then re-interpret the flow equation of the partition function in the closed channel. We also know the flow equation for the spectrum in the closed channel. This leads to the flow equation for the overlap, which gives the flow equation for the $g$-function after taking into account the exponential factor.

\paragraph{Open Channel.}

Recall that the flow equation of the spectrum in the open channel is given by
\begin{align}
\label{eq:flowEopen}
\partial_{\lambda}E_{ab}(L,\lambda)=E_{ab}(L,\lambda)\partial_L E_{ab}(L,\lambda).
\end{align}
Applying it to the open channel partition function $Z_{ab}$,
\begin{align}
Z_{ab}(R,L|\lambda)=\sum_{\psi}e^{-E_{ab}^{\psi}(L,\lambda)R},
\end{align}
we find
\begin{align}
\partial_{\lambda}Z_{ab}=&\,\sum_{\psi}R\left(-\partial_{\lambda}E_{ab}^{\psi}\right)e^{-E_{ab}^{\psi}R}\\\nonumber
=&\,\sum_{\psi}R\left(-E_{ab}^{\psi}\partial_L E_{ab}^{\psi}\right) e^{-E_{ab}^{\psi}R}\\\nonumber
=&\,-(\partial_R-R^{-1})\partial_LZ_{ab}.
\end{align}
Therefore, we have the following flow equation
\begin{align}
\partial_{\lambda}Z_{ab}(R,L|\lambda)=-\left(\frac{\partial}{\partial R}-\frac{1}{R}\right)\partial_L Z_{ab}(R,L|\lambda).
\end{align}
This flow equation has been obtained first by Cardy in \cite{Cardy:2018sdv} from random geometry considerations, which serves as a consistency check for our result.
We can also apply the flow equation to the closed channel partition function. We define $F_{ab}^{\phi}(R)=(G_a^{\phi}(R))^*G_b^{\phi}(R)$, identifying each Boltzmann weight, we have
\begin{align}
\partial_{\lambda}\left(F_{ab}^{\phi}(R,\lambda)e^{-E_{\phi}(R,\lambda)L}\right)=-\left(\frac{\partial}{\partial R}-\frac{1}{R}\right)\partial_L\left(F_{ab}^{\phi}(R,\lambda)e^{-E_{\phi}(R,\lambda)L} \right).
\end{align}
Expanding both sides, we find that the $\mathcal{O}(L)$ piece reduces to the flow equation of the deformed spectrum $E_{\phi}(R,\lambda)$ in the closed channel
\begin{align} \label{eqn-FlowER}
\partial_{\lambda}E_{\phi}(R,\lambda)=E_{\phi}(R,\lambda)\partial_R E_{\phi}(R,\lambda),
\end{align}
while the $\mathcal{O}(1)$ piece gives the flow equation for $F_{ab}^{\phi}$
\begin{align} \label{eqn-FlowFab}
\partial_{\lambda}F_{ab}^{\phi}(R,\lambda)=\left(\frac{\partial}{\partial R}-\frac{1}{R}\right)(F_{ab}^{\phi}\,E_\phi)\, .
\end{align}
Now we specify to the ground state and denote the corresponding quantity as $F_{ab}$. The product of $g$-functions is given by
\begin{align}
g_{ab}=g_a^* g_b=e^{(f_a(\lambda)+f_b(\lambda))R}F_{ab}.
\end{align}
To write down the flow equation for $g_{ab}$, we need to know the flow equation for $f_a(\lambda)+f_b(\lambda)$. This can be derived by considering the large $L$ limit of the flow equation (\ref{eq:flowEopen}).
By comparing the coefficients of $L$, we find that
\begin{align}
\label{eq:flowfab}
\epsilon'_0(\lambda)=\epsilon_0(\lambda)^2,\qquad f'_a(\lambda)+f'_b(\lambda)=\epsilon_0(\lambda)(f_a(\lambda)+f_b(\lambda)).
\end{align}
These equations can be solved, which leads to
\begin{align}
\epsilon_0(\lambda)=\frac{\epsilon_0}{1-\lambda\epsilon_0},\qquad f_a(\lambda)+f_b(\lambda)=\frac{f_a+f_b}{1-\lambda\epsilon_0},
\end{align}
where the quantities on the right hand side of the above equations are undeformed. We can then derive the flow equation for the quantity $g_{ab}(R,\lambda)$ as
\begin{align}
\partial_{\lambda}g_{ab}=&\,R(f'_a(\lambda)+f'_b(\lambda))g_{ab}+e^{(f_a+f_b)R}\partial_{\lambda}F_{ab}\\\nonumber
=&\,R \epsilon_0(f_a(\lambda)+f_b(\lambda))g_{ab}+e^{(f_a+f_b)R}\partial_R \Big(F_{ab} E_\phi \Big)-\frac{1}{R}g_{ab} E_\phi.
\end{align}
Using the fact that
\begin{align}
e^{(f_a+f_b)R}\partial_RF_{ab}=\partial_R g_{ab}-(f_a+f_b)g_{ab},
\end{align}
and the flow equation (\ref{eq:flowfab}), we obtain
\begin{align} \label{eqn-Flowgab}
    \partial_{\lambda}g_{ab}=(R\epsilon_0-E_\phi)(f_a+f_b)g_{ab}+(\partial_R-R^{-1})\big(g_{ab} E_\phi\big).
\end{align}

\paragraph*{Example: Free Theory and CFTs.}

We can easily verify the flow equation \eqref{eqn-FlowFab}, or its equivalent form \eqref{eqn-Flowgab} for free theories and CFTs, where the exact $g$-function is known \cite{Cavaglia:2016oda}:
\beq
G_a^\phi(R,\lambda) = G_a^{\phi, \textrm{CFT}} \sqrt{\frac{R}{R+\lambda E_\phi(R,\lambda)}}\, .
\eeq
Here $G_a^{\phi, \textrm{CFT}}$ is a constant that only depends on the boundary condition, but not the size of the system. It implies that we could treat it as a constant, when studying the flow equation.
Comparing with the definition of $F_{ab}$, we immediately see that
\beq
F_{ab}^\phi = (G_a^\phi)^* G_b^\phi = (G_a^{\phi, \textrm{CFT}})^* G_b^{\phi, \textrm{CFT}} \frac{R}{R+\lambda E_\phi(R,\lambda)}.
\eeq
Using the flow equation of the energy \eqref{eqn-FlowER}, one immediately finds the deformed $G_a^\phi$ defined above solves the flow equation \eqref{eqn-FlowFab}.

\paragraph{General Solutions.}

We can simplify the structure of \eqref{eqn-FlowFab}. Defining
\beq
F_{ab}^\phi = \widetilde{F}_{ab}^\phi \frac{R}{R+\lambda E_\phi(R,\lambda)},
\eeq
the flow equation of $\widetilde{F}_{ab}$ simplifies to
\beq \label{eqn-FlowGModified}
\partial_\lambda \log \widetilde{F}_{ab} = E_\phi \partial_R \log \widetilde{F}_{ab} + \partial_R E_\phi.
\eeq
Denoting $\log \widetilde{F}$ by $\mathcal{F}$, we can solve this first order partial differential equation for $\mathcal{F}$ by the method of characteristics. We refer to appendix~\ref{apd-Gfunction} for a brief review of this method. Introducing an auxiliary ``time'' parameter $t$, the equation above is equivalent to the following set of ordinary differential equations,
\beq
\frac{\dd \lambda(t)}{\dd t} = 1, \quad \frac{\dd R(t)}{\dd t} = - E_\phi(R(t),\lambda(t)), \quad \frac{\dd \mathcal{F}}{\dd t} = \partial_R E_\phi \, ,
\eeq
where we treat $\lambda, R$ and their functions as functions of $t$.

We choose the initial value of those equations in such way that when $t=0$, the solutions correspond to the undeformed theory, where $\lambda =0$. The first equation then immediately implies $\lambda = t$. For the second equation, notice that the \textit{total differential} of $E_\phi$ vanishes due to the flow equation \eqref{eqn-FlowER}:
\beq
\frac{\dd E_\phi}{\dd t} = \frac{\dd R}{\dd t} \partial_R E_\phi + \frac{\dd \lambda}{\dd t} \partial_\lambda E_\phi = \partial_\lambda E_\phi - E_\phi \partial_R E_\phi = 0.
\eeq
Therefore, the RHS of the second equation is actually independent of $t$. The solution is then obtained by integrating once,
$R(t) = R_0 - t E_0(R_0)$, where $E_0$, $R_0$ denote the undeformed energy and the undeformed circumference, respectively. In order to solve the last equation, notice that
\beq
\frac{\dd}{\dd t} (\partial_R E_\phi) = \frac{\dd R}{\dd t} \partial_R^2 E_\phi + \frac{\dd \lambda}{\dd t} \partial_\lambda \partial_R E = \partial_R \partial_\lambda E_\phi - E_\phi \partial_R^2 E_\phi = (\partial_R E_\phi)^2,
\eeq
where we take another $\partial_R$ derivative of the flow equation \eqref{eqn-FlowER} to simplify the RHS. Therefore, as a function of $t$, $\partial_R E$ is easily found to read
\beq
(\partial_R E_\phi)(t) = \frac{\partial_{R_0} E_0}{1- t \partial_{R_0} E_0},
\eeq
where the derivative $\partial_{R_0} E_0$ is evaluated at $t=0$, namely in the undeformed theory. The solution for $\mathcal{F}$ is then easily obtained,
\beq
\mathcal{F}(t, R_0) = \mathcal{F}_0(R_0) - \log \Big(1- t\, \partial_{R_0} E_0(R_0) \Big),
\eeq
where $\mathcal{F}_0 = \log F_{ab}(\lambda = 0, R_0)$ is the undeformed $g$-function.

To summarize, we have found the general solution to the flow equation $\mathcal{F} = \log F_{ab}$, which depends on two parameters $(R_0,t)$. In order to convert them into the original variables $(R,\lambda)$, we have to solve them in terms of $(R_0,t)$. Explicitly, they are given by the implicit solutions of the following equations,
\beq
t = \lambda, \quad R = R_0 - t E_0(R_0) = R_0 - t E(R(t),\lambda(t))\, .
\eeq
where $\mathcal{F}_0 = \log F_{ab}(\lambda = 0)$ is the undeformed $g$-function. Substituting the solution $t= t(R,\lambda), R_0 = R_0(R,\lambda)$ to $\mathcal{F}(t,R_0)$, we find the final solution of $\log F_{ab}$. 

For free theory, the ground state energy is given by the Casimir energy $E_0(R_0) = -\pi c/6R_0$, where $c$ is the effective central charge. In this case, the coordinate transformation between $(t,R_0)$ and $(\lambda, R)$ can be made explicit,
\beq
t = \lambda, \quad R_0 = \frac{1}{2} \Big( R + \sqrt{R^2 - \frac{2 c\pi \lambda}{3} } \Big).
\eeq
One can easily solve $\mathcal{F}$,
\beq
\mathcal{F}(\lambda,R) = \mathcal{F}_0 (R_0(\lambda,R)) - \log \Big(1- \frac{c \pi \lambda}{6R^2_0}\Big).
\eeq
Substituting the function $R_0(R,\lambda)$, one can easily verify the flow equation \eqref{eqn-FlowGModified} holds, regardless of the form of $\mathcal{F}_0$.

%%%%%%%%%%%%%%%%%%%%%%%%%%%%%%%%%%%%%%%%%%%%%%%%%%%%%%%%%%%%%%
\section{Conclusions and Outlook}
\label{sec:conclude}

While $T\bar T$-like deformations of quantum field theories have so far mainly been studied for systems with closed boundary conditions, in this paper we initiate the study of such deformations for QFTs with boundaries and defects. For this purpose we have applied the generic formalism developed in \cite{Bargheer:2008jt,Bargheer:2009xy,Loebbert:2012yd}, which allowed us to derive the deformed scattering matrices. With these quantities at hand, we derived the flow equation for the deformed finite volume spectrum for all bilocal- and odd charge- deformations. For $T\bar{T}$ deformation, we rederived the flow equation for the cylinder partition function and the deformed exact $g$-function \cite{Cardy:2018sdv}. There are plenty of directions that deserve further investigation.

First, it would be important to further explore the relation between field theory and spin chain deformations as illustrated in \tabref{tab:DifferentDeformations}. While the original $T\bar T$-deformations of  field theories have not been defined for the lattice models due to the lack of a conserved momentum charge density, the moduli space of known spin chain deformations is enlarged by the so-called \emph{boost deformations} which have no analogue in the field theory yet. It would be highly desirable to extend the respective field theory deformations to the spin chain and vice versa. Here it might be fruitful to consider the continuum limit of a specific lattice model in detail and to trace the respective deformations. Another promising direction is to take inspiration from the relation between the above boost deformations and inhomogeneous spin chains \cite{Bargheer:2009xy,Jiang:2014mja} and to investigate a field theory analogue of the latter.

Furthermore, here we have mainly focussed on deformations induced by charges of spacetime-type. Similarly, the employed formalism allows us to use charges corresponding to internal symmetries \cite{Beisert:2013voa}, in analogy to the $J\bar T$ deformations in field theory \cite{Guica:2017lia,Guica:2019vnb,Anous:2019osb,Nakayama:2018ujt,Frolov:2019xzi,Aguilera-Damia:2019tpe,Chakraborty:2019mdf,Apolo:2018qpq}. It should be enlightening to extend our analysis of boundaries and defects to the class of $J\bar T$ deformations in field theory.

Moreover, the leading order classical analysis in \secref{sec:ClassDef} suggests that for a given $T\bar T$-deformed bulk model, the boundary Hamiltonian $\theta(\phi)$ has to be trivial in order to preserve integrability. It would be desirable to prove or falsify this statement at higher orders in the deformation parameter and for generic deformed bulk models. As demonstrated, a way out is to allow the boundary Hamiltonian to depend also on derivatives of the field $\phi$. Further investigation of this type of models would also be desirable. Notably, in the spin chain case it has been shown that deformations of models with non-trivial boundary Hamiltonian can be performed using the formalism applied here  \cite{Beisert:2013voa}.

While in this work we introduced field theory deformations by extension of previous findings for lattice models, it should be very interesting to explore the full field theory moduli space in order to verify that no deformations were missed, see \cite{Doyon:2021tzy} for an interesting work in this direction. In the spin chain case such an analysis was performed for integrable models by making a general ansatz for the leading deformed charges and requiring a certain symmetry and locality in the sense of \eqref{eq:Hamloc}  \cite{Beisert:2005wv,Beisert:2008cf,Beisert:2013voa}. Demanding that these charges  commute, then resulted in constraints, whose solutions could be mapped to the boost and bilocal deformation generators discussed above. A similar analysis in the field theory context might reveal new types of deformations that have not been explored so far.

Finally, in the present paper we have put an emphasis on systems with integrable boundaries. While this is convenient, there is a priori no reason to assume integrability in the field theory context, where deformations merely require the conserved momentum and Hamiltonian. Investigating explicit examples of boundary theories beyond the scope of integrability will be highly interesting and should be feasible with the  methods employed here.

%%%%%%%%%%%%%%%%%%%%%%%%%%%%%%%%%%%%%%%%%%%%%%%%%%%%%%%%%%%%%%

\subsection*{Acknowledgements}

We thank the organizers of the workshop \emph{TTbar Deformation and Integrability} at APCPT in 2020 where the project was initiated. The work of DlZ was supported in part by a center of excellence funded by the Israel Science Foundation (grant number 2289/18) and by the Israel Science Foundation (grant number 1197/20). DlZ is grateful to CERN for the warm hospitality in the early stage of this project. The work of FL is funded by the Deutsche Forschungsgemeinschaft (DFG, German Research Foundation)--Projektnummer 363895012.

\appendix

\section{Classical Deformation of the Sine-Gordon Model}
\label{apd-SG}

Consider the Sine-Gordon (SG) model. Its action is given by
\beq
S = \int d^2 x \Big( \frac{1}{2} (\partial_\mu \phi)^2 - \frac{m^2}{\beta^2} \cos \beta \phi \Big) + \int dy\, \theta(\phi) \, .
\eeq
Based on classical analysis, in \cite{Ghoshal:1993tm} the authors have determined the most general boundary potential $\theta(\phi)$ that is compatible with integrability
\beq
\theta(\phi) = - M \cos \frac{\beta}{2} (\phi -\phi_0),
\eeq
where $M,\phi_0$ are two free parameters.

\paragraph*{Equation of Motion.}

In the bulk, the equation of motion is obtained by taking the functional variation w.r.t $\phi$,
\beq
\Box \phi = \frac{m^2}{\beta} \sin \beta \phi \Leftrightarrow \partial \bar{\partial} \phi = \frac{m^2}{4\beta} \sin \beta \phi \, .
\eeq

\paragraph*{$T\bar{T}$ Deformation.}

Let us first compute the stress tensor. By definition,
\beq \label{eqn-StressTensor}
T_{\mu \nu}^\textrm{bulk} = \frac{\partial \mathcal{L}^\textrm{bulk}}{\partial (\partial^\mu \phi)} \partial_\nu \phi - \delta_{\mu \nu} \mathcal{L}^\textrm{bulk} =\partial_\mu \phi \partial_\nu \phi - \eta_{\mu \nu} \mathcal{L}^\textrm{bulk} \, .
\eeq
In complex coordinates we have
\begin{equation}
\begin{aligned}
T & = - T_{zz} = -(\partial \phi)^2, \quad \bar{T} = - T_{\bar{z}\bar{z}} = -(\bar{\partial} \phi)^2,  \\
\Theta & = T_{z\bar{z}} = \bar{\Theta} = T_{\bar{z}z} = \frac{m^2}{2\beta^2} \cos \beta \phi \, .
\end{aligned}
\end{equation}
It is straightforward to verify that $\bar{\partial}  T = \partial \Theta$.
Using the expressions above, we immediately find
\beq
\det T^\textrm{bulk}_{\mu \nu} = 4(T\bar{T} - \Theta \bar{\Theta}) \, .
\eeq
Therefore, if we introduce this deformation, at order $\mathcal{O}(\lambda)$ we have
\beq
S^\lambda_\textrm{bulk} = S^{\lambda =0}_\textrm{bulk} + \lambda \int d^2 x\ 4(T\bar{T} - \Theta \bar{\Theta})+ \mathcal{O}(\lambda^2) \, .
\eeq

\paragraph*{Deformed Stress Tensor.}

The deformed bulk stress tensor can be obtained using \eqref{eqn-StressTensor}:
\begin{equation}
\begin{aligned}
T & = - T_{zz} = -\frac{1}{2}\frac{\partial \mathcal{L}^\textrm{bulk}}{\partial (\bar{\partial} \phi)} \partial \phi = - (\partial \phi)^2 - 4\lambda (\partial \phi)^3 (\bar{\partial} \phi), \\
\bar{T} & = - T_{\bar{z}\bar{z}} = -\frac{1}{2}\frac{\partial \mathcal{L}^\textrm{bulk}}{\partial (\partial \phi)} \bar{\partial} \phi = - (\bar{\partial} \phi)^2 - 4\lambda (\partial \phi) (\bar{\partial} \phi)^3, \\
\Theta & = T_{z\bar{z}} = + \frac{1}{2}\frac{\partial \mathcal{L}^\textrm{bulk}}{\partial (\bar{\partial} \phi)} \bar{\partial} \phi - \frac{1}{2} \mathcal{L}_\textrm{bulk}^{(\lambda)} = \frac{m^2}{2\beta^2} \cos \beta \phi - \lambda \frac{m^2}{\beta^2} \cos^2 \beta \phi + 2\lambda (\partial \phi)^2 (\bar{\partial} \phi)^2, \\
\bar{\Theta} & = T_{\bar{z}z} = + \frac{1}{2}\frac{\partial \mathcal{L}^\textrm{bulk}}{\partial (\partial \phi)} \partial \phi - \frac{1}{2} \mathcal{L}_\textrm{bulk}^{(\lambda)} = \frac{m^2}{2\beta^2} \cos \beta \phi - \lambda \frac{m^2}{\beta^2} \cos^2 \beta \phi + 2\lambda (\partial \phi)^2 (\bar{\partial} \phi)^2\, .
\end{aligned}
\end{equation}
Therefore we find
\begin{equation}
\begin{aligned}
-\ii \left(T -\bar{T} +\bar{\Theta} -\Theta \right) \Big|_{x=0} & = \partial_x \phi \partial_y \phi \Big(1 + \lambda [(\partial_x \phi)^2 + (\partial_y \phi)^2 ] \Big).
\end{aligned}
\end{equation}
We see that if the deformed boundary Lagrangian is of potential type, this term is always be a total $y$-derivative. Next we investigate higher conserved charges.

\paragraph*{Deformed Equations of Motion.}

It is straightforward to verify that the modified equations of motion are given by
\begin{equation}
\begin{aligned}
\mathrm{Bulk}: & & 4\partial \bar{\partial} \phi (1+8 \lambda \partial \phi \bar{\partial} \phi) & = \frac{m^2}{\beta} \sin \beta \phi + \lambda \frac{m^4}{\beta^3} \sin (2\beta\phi) - 8 \lambda \Big( \bar{\partial}^2 \phi (\partial \phi)^2 + {\partial}^2 \phi (\bar{\partial} \phi)^2  \Big), \\
\mathrm{Bdr}:  & & \partial_x \phi \Big|_{x=0} & = -\theta'(\phi) -\lambda \partial_x \phi \left[ (\partial_x \phi)^2 +(\partial_y \phi)^2) \right],
\end{aligned}
\end{equation}
where we use $4\partial \phi \bar{\partial} \phi = (\partial_x \phi)^2 +(\partial_y \phi)^2 $. At order $\mathcal{O}(\lambda)$, we can solve $\partial \bar{\partial} \phi$ for the bulk equation of motion:
\beq \label{eqn-SG-bulkeom}
4\partial \bar{\partial} \phi = \frac{m^2}{\beta} \sin \beta \phi (1-8\lambda \partial \phi \bar{\partial} \phi) + \lambda \frac{m^4}{\beta^3} \sin (2\beta\phi) - 8 \lambda \Big( \bar{\partial}^2 \phi (\partial \phi)^2 + {\partial}^2 \phi (\bar{\partial} \phi)^2  \Big) + \mathcal{O}(\lambda^2).
\eeq

\paragraph*{Odd Charge: $T_3$.}

Before discussing the deformation of the first even higher charge, let us briefly discuss why there is no odd charge in SG theory. The reason is that all odd charges are total derivatives.
For instance, let us consider $T_3$. The most general form containing three derivatives would is
\beq
T_3 = \frac{1}{3} \alpha_0 (\partial \phi)^3 + \alpha_1 \partial^2 \phi \partial \phi + \alpha_2 \partial^3 \phi, \quad \Theta_1 = F(\phi) \partial \phi \, .
\eeq
Computing derivatives, we find (for simplicity, we rescale the fields and take $m^2=4$)
\begin{equation}
\begin{aligned}
\bar{\partial} T_3 & =(\partial \phi)^2 (\partial \bar{\partial} \phi) + \alpha_1 \partial(\partial \bar{\partial} \phi) \partial \phi+ \alpha_1 \partial^2 \phi (\partial \bar{\partial} \phi) + \alpha_2 \partial^2 (\partial \bar{\partial} \phi) \\
& =  (\partial \phi)^2 (\alpha_0 \sin \phi + \alpha_1 \cos \phi -\alpha_2 \sin \phi) + \partial^2 \phi (\alpha_1 \sin \phi + \alpha_2 \cos \phi), \\
\partial \Theta_1 & = F(\phi) \partial^2 \phi + F'(\phi) (\partial \phi)^2 \, .
\end{aligned}
\end{equation}
Comparing the derivatives we find that we must have $\alpha_0 =0$. What remains are just total $z$ derivatives.

\paragraph*{Deformation of $T_4$.}

The first set of non-trivial higher conserved charges is given by
$T_4$ and its conjugate:
\begin{equation}
\begin{aligned}
T_4 & = (\partial^2 \phi)^2 - \frac{\beta^2}{4} (\partial \phi)^4, \quad \Theta_2 = \frac{m^2}{4} (\partial \phi)^2 \cos \beta \phi \, , \\
\bar{T}_4 & = (\bar{\partial}^2 \phi)^2 -  \frac{\beta^2}{4} (\bar{\partial} \phi)^4, \quad \bar{\Theta}_2 = \frac{m^2}{4} (\bar{\partial} \phi)^2 \cos \beta \phi \, .\\
\end{aligned}
\end{equation}
We would like to study their one-loop deformations. To begin with, let us denote the correction term with a superscript,
\beq
T_4 \rightarrow T_4^{(0)} + \lambda T_4^{(1)} + \mathcal{O}(\lambda^2), \quad \text{etc} \, .
\eeq
We would like to choose the $T_4^{(1)}$ term to make the conservation equations work at this order.
We compute
\begin{equation}
\begin{aligned}
\bar{\partial}T_4 & = 2 (\partial^2 \phi) (\partial^2 \bar{\partial} \phi) - \beta^2 (\partial \phi)^3 (\partial \bar{\partial} \phi) + \lambda \bar{\partial} T_4^{(1)} + \mathcal{O}(\lambda^2), \\
\partial \Theta_2 & = \frac{m^2}{2} \cos \beta \phi (\partial^2 \phi) \partial \phi-\frac{m^2 \beta}{4} \sin (\beta  \phi) (\partial \phi )^2 + \partial \Theta_2^{(1)}+ \mathcal{O}(\lambda^2).
\end{aligned}
\end{equation}
We continue by substituting the bulk equation of motion \eqref{eqn-SG-bulkeom}. The $O(\lambda)$ contribution of $\bar{\partial} T_4$ is simply,
\beq
\bar{\partial}T_4 \Big|_{\mathcal{O}(\lambda)} =2 \partial^2 \phi\, \partial \Big( \partial \bar{\partial} \phi \Big|_{\mathcal{O}(\lambda)} \Big) -\beta^2 (\partial \phi)^3 \Big(\partial \bar{\partial} \phi \Big|_{\mathcal{O}(\lambda)} \Big) + \bar{\partial} T_4^{(1)}\, .
\eeq
Similarly, the $O(\lambda)$ contribution of $\partial \Theta_2$ is simply $\partial \Theta_2^{(1)}$, since the tree-level pieces does not involve any equations of motion.
After some manipulations, one finds,
\begin{equation}
\begin{aligned}
T_4^{(1)} & = 8 (\partial^2 \phi)^2 \partial \phi \bar{\partial} \phi -2 \beta^2 (\partial \phi)^5 \bar{\partial} \phi -\frac{3 m^2 (\partial^2\phi)^2 \cos (\beta  \phi)}{\beta ^2}+\frac{m^2 \partial^2 \phi (\partial \phi)^2 \sin (\beta  \phi )}{\beta }\\\nonumber
&-\frac{3}{4} m^2 (\partial \phi)^4 \cos (\beta  \phi ), \\
\Theta_2^{(1)} & =\frac{\beta^2}{2}  (\partial \phi)^4 (\bar{\partial} \phi)^2 -2 (\partial^2 \phi)^2 (\bar{\partial} \phi)^2 -\frac{m^4}{4\beta^2} (\partial \phi)^2 \big(1 + 2 \sin^2 \beta \phi \big) \, .
\end{aligned}
\end{equation}
As usual, $\bar{T}_4, \bar{\Theta}_2$ are obtained by swapping $z,\bar{z}$.

\paragraph*{Integrability-Preserving Boundary Potential.}

As for the free scalars, we expand the boundary potential to $\mathcal{O}(\lambda)$:
\beq
\theta_\lambda = \theta_{(0)} + \lambda \theta_{(1)} + \mathcal{O}(\lambda^2).
\eeq
The strategy to proceed is to eliminate two or more $x$-derivative terms by the bulk EOM, while to eliminate one $x$-derivative terms by the boundary EOM. After long but straightforward algebra, we finally find the following:

For the one-loop term, it has the following structure
\begin{equation}
\begin{aligned}
& -8\ii\left(T_4-\bar{T}_4 +\bar{\Theta}_2 -\Theta_2 \right) \Big|_{\mathcal{O}(\lambda)} \\
= & (\textrm{functions of } \phi) \partial_y \phi + \mathcal{A}_{1,1,1}(\phi) (\partial_y \phi)^3 + \mathcal{A}_{1,2}(\phi) (\partial_y \phi) \partial_y^2 \phi   \\
& +\mathcal{A}_{1,1,1,1,1}(\phi) (\partial_y \phi)^5 + \mathcal{A}_{1,1,1,2} (\partial_y \phi)^3 \partial_y^2 \phi + \mathcal{A}_{1,2,2}(\phi) (\partial_y \phi) (\partial_y^2 \phi)^2.
\\
&
\end{aligned}
\end{equation}
Here the subscript of the coefficient functions represents the accompanying derivative structure, in an apparent way. The explicit expressions of them are listed below:
\begin{equation}
\begin{aligned}
\mathcal{A}_{1,1,1}(\phi) & = \beta ^2 \theta _1'(\phi )-\frac{1}{32} \beta  M \sin \frac{\beta  \phi }{2} \left(\cos (\beta  \phi ) \left(9 \beta ^4 M^2-48 m^2\right)+32 m^2+3 \beta ^4 M^2\right),\\
\mathcal{A}_{1,2}(\phi) & =\frac{M }{8} \cos \frac{\beta  \phi }{2} \left(\cos (\beta  \phi ) \left(80 m^2-3 \beta ^4 M^2\right)-32 m^2+3 \beta ^4 M^2\right)-8 \theta _1''(\phi ), \\
\mathcal{A}_{1,1,1,1,1}(\phi) & = +\frac{3}{8} \beta ^3 M \sin \frac{\beta  \phi }{2},\\
\mathcal{A}_{1,1,1,2}(\phi) & = -3 \beta ^2 M \cos \frac{\beta  \phi }{2},\\
\mathcal{A}_{1,2,2}(\phi) & =+ 6 \beta  M \sin \frac{\beta  \phi }{2}\, .\\
\end{aligned}
\end{equation}
Notice that all five derivative terms are \textit{independent} of $\theta_1$, so they must combine to be a total derivative by themselves.

Since the structure of the five-derivative terms missed a structure that is proportional to $(\partial_y \phi)^2 (\partial_y^3\phi)$, it cannot be written as total derivatives. Thus, the only solution is to take $M =0$. If it is the case,
\begin{equation}
-\ii\left(T_4^{(1)}-\bar{T}_4^{(1)} +\bar{\Theta}_2 -\Theta_2 \right)
= \lambda  \left(\beta ^2 \theta_1' \left(\frac{\partial \phi}{\partial y}\right)^3-8 \frac{\partial \phi}{\partial y} \frac{\partial ^2\phi}{\partial y^2} \theta_1''\right)
+ \text{total $y$-derivatives} .
\end{equation}
For this term to be a total derivative, rewriting the first term as
\beq
\theta_1' \left(\frac{\partial \phi}{\partial y}\right)^3= \frac{d}{dy} \Big(\theta_1 (\partial_y \phi)^2 \Big) - 2\theta_1 \partial_y \phi \partial^2_y \phi \, .
\eeq
Therefore, a necessary condition for being a total derivative is
\beq
\beta^2 \theta_1 + 4 \theta_1'' =0 \Rightarrow \theta_1 = M_1 \cos \frac{\beta (\phi - \phi_1)}{2} \, ,
\eeq
with $M_1, \phi_1$ constant of integrations.

The conclusion is, at $\mathcal{O}(\lambda)$, the most general boundary potential stays the same form, but it must be \textit{delayed} for one order. It suggests that the only compatible boundary potential with bulk $T\bar{T}$ deformation is zero, as for the free case.

\section{Generalized Partition Function and Higher Deformations}
\label{apd-HigherDeform}

\paragraph{Generalized Partition Function.}

Unlike non-integrable systems, which are fully characterized by the Boltzmann weight at thermal equilibrium, the integrable systems do not fully thermalize. Explicitly, it means that the thermal eigenstates carry additional quantum numbers that are associated with the infinite set of higher conserved charges. This will lead to technical difficulties.
\paragraph{Toy Model of Higher Deformation.}

In this appendix, we shall illustrate the technical difficulties of deriving a flow equation for the generalized partition function.

Let us consider the following toy model, whose partition function only contains one higher charge $Q$,
\begin{equation}
    Z = \sum_n e^{-\beta E_n -\mu Q_n},
\end{equation}
where we expand the charges on the complete eigenbasis of the Hamiltonian.

Suppose the system is deformed by the higher charge $Q$, explicitly, the energy and the charges will be deformed as
\begin{equation}
    \partial_\lambda E_n = Q_n \partial_\nu E_n, \quad \partial_\lambda Q_n = Q_n \partial_\nu Q_n.
\end{equation}
Here $\nu$ represents the twist.

Applying the flow equation to the partition function, we find
\beq
\partial_\lambda Z = - \sum_n (\beta \partial_\lambda E_n + \mu \partial_\lambda Q_n) e^{-\beta E_n -\mu Q_n} = - \sum_n Q_n (\beta \partial_\nu E_n + \mu \partial_\nu Q_n) e^{-\beta E_n -\mu Q_n}.
\eeq
Compare with the $\partial_\nu$ derivative of the partition function,
\begin{equation}
    \partial_\nu Z = - \sum_n (\beta \partial_\nu E_n + \mu \partial_\nu Q_n) e^{-\beta E_n -\mu Q_n},
\end{equation}
one immediately sees that
\beq
\partial_{\mu} \partial_\nu Z = -\partial_\lambda Z - \sum_n \partial_\nu Q_n e^{-\beta E_n -\mu Q_n}.
\eeq
Unfortunately, it seems hard to rewrite the second term on the RHS of the previous equation as a differential operator acting on the partition function like in the $T\bar{T}$ case. This is the barrier for us to write down a simple flow equation for the deformed partition function. This difficulty is likely to be related to the fact that we do not have a gravity description of this type of deformation. It is interesting to compare the current situation with the one of $JT_a$ deformation. There, we also have an additional $U(1)$ current and the corresponding charge $Q$ enters the generalized partition function. However, the important difference is that in that the $U(1)$ conserved charge \emph{does not} depend on rapidities. Therefore it does not flow under the deformation. In our situation, the higher conserved charge are rapidity dependent, which also flows under the deformation. Suppose we can write down a simple flow equation for the partition function. Since the partition function can be written in the Lagrangian formalism, it would be possible to derive it from the point of view of 2d gravity. Such a situation would be `too nice to be true'. The higher conserved charges might be related to coupling the theory to some higher spin theory.

%%%%%%%%%%%%%%%%%%%%%%%%%%%%%%%%%%%%%%%%%%%%%%%%%%%%%%%%%%%%%%%%%%%
\section{Method of Characteristics}
\label{apd-Gfunction}
%%%%%%%%%%%%%%%%%%%%%%%%%%%%%%%%%%%%%%%%%%%%%%%%%%%%%%%%%%%%%%%%%%%
In this review, we review the basic idea of the \emph{method of characteristics}. Consider the following boundary value problem, characterized by a partial differential equation
\beq
a(x,y) \frac{\partial u}{\partial x} + b(x,y) \frac{\partial u}{\partial y} =c(x,y), \qquad u(x,y=0) = f(x)\, ,
\eeq
where for simplicity we assume the variables are $x,y$, and the unknown function is denoted by $u(x,y)$.  $a,b,c$ and $f$ are known functions.

The crucial idea of the method of characteristics is to treat the equation above as a geometric constraint. Explicitly, the equation above is equivalent to the condition that the vector field $(a,b,c) \in \mathbb{R}^3$ is always perpendicular to the normal of the surface $z=u(x,y)$. (One can think of $u$ as taking values on the $z$ axis.) If we start at an arbitrary point on the surface, this implies that all points on the \textit{integral curve} of the vector field $(a,b,c)$ will remain on the surface. If the integral curves exist globally and do not intersect, regardless of the initial points, the surface $u(x,y)$ is just the union of all integral curves. In our case, this means that the solution $u$ can by obtained by studying all the integral curves of $(a,b,c)$ that start at the boundary, where $y=0$.

To be more precise, let us first pick an arbitrary point $(x_0,y=0)$ on the boundary surface. The integral curve of $(a,b,c)$ with such an initial condition can then be uniquely determined by solving the following set of ordinary differential equations,
\beq
\frac{\dd x(s)}{\dd s} = a(x(s),y(s)), \quad \frac{\dd y(s)}{\dd s} = b(x(s), y(s)),
\quad \frac{\dd u(s)}{\dd s} = c(x(s), y(s)),
\eeq
where we parametrize the curve by $s$. In the literature, the differential equations for $x$ and $y$ are called the \textit{characteristic equations},  while the last equation for the unknown function $u$ follows from the first two equations.

Solving the characteristic equations and fixing the parametrization of the integral curve by demanding $x(s=0) = x_0, y(s=0) = 0$, we obtain the explicit parametrization
\beq
x(s) = f_1(x_0, s), \quad y(s) = f_2(x_0, s),
\eeq
on the integral curve. Plugging them into the third differential equation for $u(s)$, we find a unique solution $u(s,x_0)$, satisfying $u(s=0) = u(x_0,y=0) = f(x_0)$.

What does this have to do with the original PDE? The point is that for a generic point $(x_1, y_1)$, we can construct an integral curve that passes through it. Suppose this curve starts at $(x_0,0)$ and passes through $(x_1,y_1)$ at a time $s_0$, then the solution $u(s_0,x_0)$ automatically solves the PDE. The only thing that remains is to do a coordinate transformation that maps $(x_0, s_0)$ to the original Cartesian coordinates $(x_1, y_1)$.

To summarize, the method of characteristics consists of three steps:
\begin{enumerate}
    \item Pick an arbitrary point $(x_0,y=0)$ on the boundary and solve the characteristic equations
    $\dd x/\dd s = a, \dd y/\dd s = b$. The ambiguity of the parametrization is eliminated by demanding that the boundary point $(x_0,0)$ corresponds to $s = 0$. This will provide a coordinate transformation $(x,y) \mapsto (x_0, s)$.
    \item Using known solutions for $x,y$, we can solve for $u$ as a function of $(x_0,s)$ as well. The initial value of $u$, is nothing but $f(x_0)$. In other words, for fixed $x_0$, the integral curve gives a \textit{line} of solutions to the original PDE.
    \item We now have the solution $u(x_0, s)$. By \textit{changing} $x_0$ we can reach any point $(x_1, y_1)$ at finite time $s_0$. One can then take a coordinate transformation, $(x_0, s_0) \mapsto (x_1, y_1)$ by solving
    \beq
    x_1 = x(x_0, s_0), \quad y_1 = y(y_0, s_0)\, .
    \eeq
    The solution $u(x_1,y_1)$, for arbitrary $(x_1,y_1)$ is then obtained by substituting $(x_0, s_0)$ in the solution $u(x_0, s_0)$.
\end{enumerate}

%%%%%%%%%%%%%%%%%%%%%
%%%%%%%%%%%%%%%%%%%%%%%%%%%%%%%%%%%%%%%%%%%%%%%%%%%%%%%%%%%%
\pdfbookmark[1]{\refname}{references}
\bibliographystyle{nb}
\bibliography{BoundaryTTbarQuantum}

\end{document}